\RequirePackage{amsmath}
\documentclass[cmp]{svjour}
\usepackage{latexsym}
\usepackage{graphics}
\usepackage[utf8]{inputenc}
\usepackage{multirow}
\usepackage{diagbox}
\usepackage{amssymb,setspace}
\usepackage{mathabx} 
\usepackage{parskip}
\usepackage{subfig}
\usepackage{graphicx}

\usepackage{vmargin}
\usepackage{color}
\date{\today}

\newcommand{\ff}[5]{
\setlength{\arraycolsep}{1pt}
\renewcommand{\arraystretch}{0.8}
{f}\!\left(\!
\begin{array}{rcl}
 &#1&  \\
 #2 & #3 & #4 \\
 & #5 & 
\end{array}\!
\right)
}

\newcommand{\fff}[7]{
\setlength{\arraycolsep}{1pt}
\renewcommand{\arraystretch}{0.8}
{f}\!\left(\!
\begin{array}{rcl}
 &#1& {\color{blue}#7} \\
 #2 & #3 & #4 \\
{\color{blue}#6} & #5 &  
\end{array}\!
\right)
}

\newcommand{\dimer}[2]{
D_{\substack{#1\\#2}}
}

\newcommand{\vv}[2]{
{{\bf #1}_{#2}}
}

\newcommand{\X}{\widetilde{X}}
\newcommand{\C}{\mathcal{C}}
\newcommand{\D}{\mathcal{D}}
\newcommand{\SF}{\mathcal{S}}
\newcommand{\Per}{\mathcal{P}}
\newcommand{\NN}{\mathcal{N}}

\newcommand{\bx}{\mathbf{x}}
\newcommand{\bze}{\mathbf{0}}
\newcommand{\bi}{\mathbf{i}}
\newcommand{\bj}{\mathbf{j}}

\newcommand{\qm}{q^*}
\newcommand{\Z}{\mathbb{Z}}
\newcommand{\N}{\mathbb{N}}
\newcommand{\R}{\mathbb{R}}
\newcommand{\PP}{\mathbf{\Omega}}
\newcommand{\CP}{\mathbf{\Lambda}}

\newtheorem{thm}{Theorem}[section]
\newtheorem{defn}[thm]{Definition}
\newtheorem{lem}[thm]{Lemma}

\newtheorem{rem}[thm]{Remark}
\newtheorem{con}[thm]{Conjecture}

\usepackage{letltxmacro}
\LetLtxMacro{\originaleqref}{\eqref}
\renewcommand{\eqref}{Eq.\ \originaleqref}

\begin{document}
\title{A split-and-perturb decomposition 
of number-conserving cellular automata}


\author{Barbara Wolnik\inst{1} \and Anna Nenca\inst{2} \and Jan~M.~Baetens\inst{3} \and Bernard De Baets\inst{3} %
}                     
\institute{Institute of Mathematics, Faculty of Mathematics, Physics and Informatics, University of Gda\'{n}sk, 80-308 Gda\'{n}sk, Poland %
\and %
Institute of Informatics, Faculty of Mathematics, Physics and Informatics, University of Gda\'{n}sk, 80-308 Gda\'{n}sk, Poland%
\and %
KERMIT, Department of Data Analysis and Mathematical Modelling, Faculty of Bioscience Engineering, Ghent University, Coupure links 653, B-9000 Gent, Belgium%
}
\date{Received: date / Accepted: date}
%
\communicated{name}
\maketitle
\begin{abstract}
This paper concerns $d$-dimensional cellular automata with the von Neumann neighborhood that conserve the sum of the states of all their cells. 
These automata, called number-conserving or density-conserving cellular automata, are of particular interest to mathematicians, computer scientists and physicists, as they can serve as models of physical phenomena obeying some conservation law. 
We propose a new approach to  study such cellular automata that works in any dimension $d$ and for any set of states $Q$. Essentially, the local rule of a cellular automaton is decomposed into two parts: a split function and a perturbation. This decomposition is unique and, moreover, the set of all possible split functions has a very simple structure, while the set of all perturbations forms a linear space and is therefore very easy to describe in terms of its basis. 
We show how this approach allows to find all number-conserving cellular automata in many cases of $d$ and $Q$. In particular, we find all three-dimensional number-conserving CAs with three states, which until now was beyond the capabilities of computers. 
\end{abstract}

\maketitle

\section{Introduction}
Since cellular automata (CAs) reflect the assumption that all laws (physical, sociological, economic and so on) must result from interactions that are strictly local, they are highly suitable as discrete dynamical models of various complex phenomena. Not surprisingly then, CAs are of great interest to researchers in a broad range of scientific disciplines. For example,  CAs have recently found applications in disciplines as diverse as biology~\cite{BOWNESS201887,Nava2017}, environmental sciences~\cite{BOUAINE201836,NAGATANI2018803}, materials science~\cite{BAKHTIARI20181,YANG2018281}, pedestrian dynamics~\cite{FU201837,Qiang2018}, urban transport~\cite{IWAN2018104,WU201869}, hydrology~\cite{HYDROLOGY} and agriculture~\cite{ZHANG2018248}, to name but a few.

In recent years, scientists are more and more apt to use multidimensional or multi-state CAs.
Unfortunately, with the increase in dimension and/or the increase in the number of states, the set of all CAs 
grows rapidly. As a consequence, conventional methods, such as scanning through the entire set to find the 
CAs one is interested in, are no longer applicable. Hence, developing new tools for multidimensional/multi-state 
CAs is of the utmost importance.

The goal of this paper is thus to develop methods to study $d$-dimensional cellular automata with the von Neumann neighborhood,  \emph{i.e.}, CAs that are updating the states of the cells on the basis of the states of adjacent cells only. 
In view of incorporating conservation laws, a key requirement in physics, many models are based on a particular 
type of CAs, namely those that have the special feature of preserving the sum of the states upon every update
of all cells. Such CAs, called number-conserving CAs or density-conserving CAs, when non-integer states are allowed, were introduced by Nagel and Schreckenberg~\cite{NS} in the early nineties.
Number-conserving CAs have received ample attention, especially as models of systems of interacting particles moving in a lattice \cite{7818615,MOREIRA2004285}. In particular, such CAs appear naturally in the context of gas or fluid flow~\cite{PhysRevLett.56.1505,PhysRevA.13.1949}, and highway traffic~\cite{Belitsky2001,KKW02,PhysRevLett.90.088701,XIANG2018}.  Our focus in this paper is on this important class of CAs.

The von Neumann neighborhood is a natural choice when modelling physical phenomena. Unfortunately, studying multidimensional CAs with this kind of neighborhood is very complicated, because it is not a Cartesian product of one-dimensional neighborhoods (in contrast to the Moore neighborhood). For this reason, the problem of number conservation in $d$-dimensional CAs with the von Neumann neighborhood has been poorly investigated for $d>1$.

Obviously, for a given CA, one does not need new tools to determine whether it is number-conserving or not. In one dimension, necessary and sufficient conditions for a CA to be number-conserving were given by Boccara and Fukś~\cite{BoccaraF02}, and similarly for two or more dimensions by Durand et al.~\cite{Durand2003}. In the latter work the Moore neighborhood is considered, so  the results can be used for the von Neumann neighborhood as well. However, if one wants to find all number-conserving CAs for a~given $d$ and $Q$ it is, in general, impossible to check all CAs to find the number-conserving ones, due to the huge cardinality of the search space. In particular, it is not advisable to consider the von Neumann neighborhood as a subset of the Moore neighborhood as the first one has only $2d+1$ cells, while the second one has as many as $3^d$ cells.

The first characterization of two-dimensional number-conserving CAs with the von Neumann neighborhood was obtained by Tanimoto and Imai~\cite{TI}. Their result is stated in terms of so-called \emph{flow functions} (in the vertical, horizontal and diagonal direction) and allows to create two-dimensional number-conserving CAs. Unfortunately, it is still not of much use to find all two-dimensional number-conserving CAs, even in the case of the state set $Q=\{0,1,2\}$. However, using these flow functions they succeeded in describing all two-dimensional five-state number-conserving CAs with the von Neumann neighborhood that are rotation-symmetric, \emph{i.e.}, are invariant under rotation of the neighborhood by 90 degrees~\cite{Imai2015}.
The results presented in~\cite{TI,Imai2015} concern only $d = 2$, and the ideas used therein, in particular the flow functions, have not been transferred to higher dimensions. Even if we could use similar tools for $d>2$, the results would be of no practical value, while using them to find all number-conserving CAs would require computational power beyond current technical capabilities.

In~\cite{NCCA}, using a novel approach based on a geometric analysis of the structure of the von Neumann neighborhood in higher dimensions, necessary and sufficient conditions for a $d$-dimensional CA to be number-conserving are formulated in terms of the local rule in a similar way as in~\cite{BoccaraF02}. These conditions apply for any state set $Q\subset \R$, whether it is finite or not.
The main result presented in~\cite{NCCA} allows to find all two-dimensional three-state number-conserving CAs \cite{enum} and all two-dimensional six-state rotation-symmetric number-conserving CAs \cite{rotation}. Moreover, it allows to describe all affine continuous density-conserving CAs with state set $Q=[0,1]$, which is infinite~\cite{2ACCA}. 
However, the necessary and sufficient conditions presented in~\cite{NCCA} can be formulated in $(2d+1)2^{d^2}$ different forms, where $d$ is the considered dimension. Although they all are equivalent, the obtained formulas can differ in the number of terms. 
A better understanding of this fact has guided us towards a completely new approach to the study of number conservation.
 
 In this paper we lay bare that the local rule of any number-conserving CA with the von Neumann neighborhood can be decomposed into two parts. The first one is a~split function -- a special local function that acts as follows: each state splits into pieces according to its recipe, irrespective of the states of its neighbors.
The second one is a perturbation -- a~local function that for a~number-conserving local rule is the only possible derogation from being a split function. Both split functions and perturbations can take values outside the considered state set, but this is a small disadvantage compared to the benefits they bring. First of all, the decomposition is unique. 
Moreover, the set of all possible split functions has a~very simple structure, while the set of all perturbations forms a linear space and is therefore very easy to describe in terms of its basis. This decomposition of a number-conserving local rule allows us to further simplify the necessary and sufficient conditions formulated in~\cite{NCCA}: it is possible to find separate conditions on split functions and on perturbations. This greatly reduces the cardinality of the set of CAs that are potentially number-conserving and makes it possible to find all number-conserving CAs for a much broader range of state sets $Q$ in $2$, $3$ and higher dimensions. 
Furthermore, ongoing work has shown that the decomposition theorem
enables us to prove some general facts about number-conserving CAs. 
It is worth emphasizing that the main result presented in this paper, although very powerful, is obtained through the use of basic mathematical tools.

This paper is organized as follows. 
In Section 2 the basic concepts and notations are introduced. Section 3 presents the decomposition of a~ number-conserving local rule into a split function and a~perturbation. A demonstration of how much this decomposition reduces the computational complexity of finding all number-conserving CAs is shown in Section 4 by simple manual counting for several classical examples of $d$ and $Q$, which until now could only be achieved using a~computer. In addition, we consider also three-dimensional CAs and find all three-state number-conserving CAs with the von Neumann neighborhood, which was previously infeasible. Section 5 concludes the paper with open problems.

\section{Preliminaries}
\label{sec:prelim}

In this section, we introduce CAs and recall some results that we will use in the following sections. To define a~CA, one needs to specify a~space of cells, a neighborhood, a state set and a local rule. However, to develop our idea of decomposition, we additionally need to generalize local rules, in the sense that they can yield values outside the considered state set.

\subsection{The cellular space}\label{cs}

Let us fix the dimension $d\geq 1$ and natural numbers $n_1,n_2,\ldots,n_d$ greater than $4$. We consider the cellular space as a grid with periodic boundary conditions, \emph{i.e.},
\[
\C=\left(\Z / n_1\Z\right)\times \left(\Z / n_2\Z\right)\times\ldots\times \left(\Z / n_d\Z\right)
\]
\[
=\{0,1,\ldots,n_1-1\}\times\{0,1,\ldots,n_2-1\}\times\ldots \times\{0,1,\ldots,n_d-1\}\, .
\]
With this notation, each cell $\bi\in\C$ is a $d$-tuple 
$(i_1,\ldots, i_d)$, where $i_k\in\Z / n_k\Z$. If $d=2$, we denote cells by $(i,j)$ rather than $(i_1, i_2)$. 
Denoting the number of elements of a set $A$ as $|A|$, we have $|\C|=n_1\cdot n_2\cdot\ldots \cdot n_d$. 

\begin{figure}[!ht]
\centering
 \subfloat[]{
\includegraphics[height=0.30\textwidth]{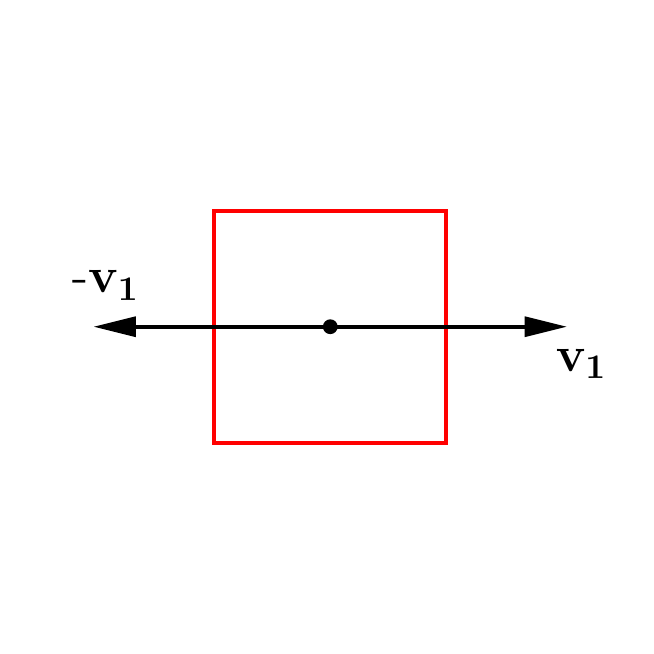}
}
 \subfloat[]{
\includegraphics[height=0.30\textwidth]{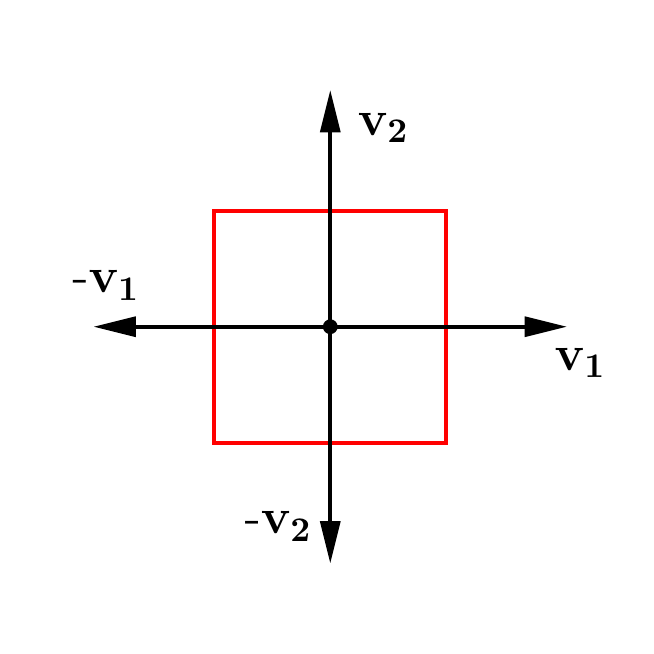}
}
 \subfloat[]{
\includegraphics[height=0.30\textwidth]{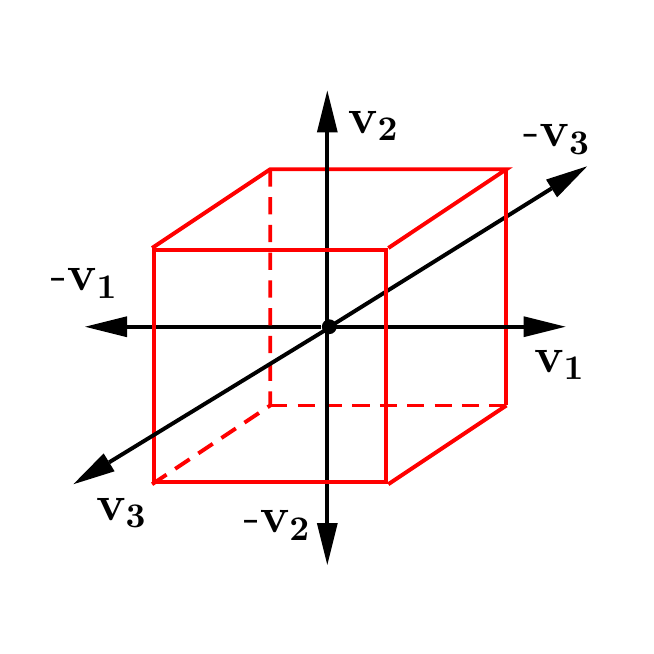}
}
\caption{Neighborhood directions in the case of one- (a), two- (b) and three- (c) dimensional CAs.}
\label{neighborhoodFig}
\end{figure}

As a consequence of the periodic boundary conditions, each cell in $\C$ has exactly $2d$ adjacent cells: two in each of the orthogonal axis directions. For example, if $d=1$, each cell has only two adjacent cells: in the direction $\vv{v}{1}$ (right) and in the direction $\mbox{-}\vv{v}{1}$ (left) (see Fig.~\ref{neighborhoodFig}(a)). If $d=2$, there are four adjacent cells: two in the horizontal directions $\vv{v}{1}$, $\mbox{-}\vv{v}{1}$ and two in the vertical directions $\vv{v}{2}$ (up) and $\mbox{-}\vv{v}{2}$ (down), as shown
in Fig.~\ref{neighborhoodFig}(b).
If $d=3$, there are two additional adjacent cells: in the direction $\vv{v}{3}$ (forward) and in the direction $\mbox{-}\vv{v}{3}$ (backward),
as shown in Fig.~\ref{neighborhoodFig}(c).

For the sake of generality, we introduce the following notation.
For each $k\in\{1,2,\ldots,d\}$, we define the vector  $\vv{v}{k}=(0,0,\ldots,0,1,0,\ldots,0)\in \R^d$, where the $k$-th component is $1$ and the other ones are zero. Let us denote the set of all considered directions as $V_+$, \emph{i.e.},
\[
V_+=\{\vv{v}{1},\mbox{-}\vv{v}{1},\vv{v}{2},\mbox{-}\vv{v}{2},\ldots, \vv{v}{d},\mbox{-}\vv{v}{d}\}\, .
\]
Additionally, let $\vv{0}{}=(0,0,\ldots,0)\in \R^d$ and
$V= V_+\cup \{\vv{0}{} \}$. Then for $\bi \in \C$ and $\vv{v}{}\in V$, $\bi +\vv{v}{}\in \C$ is nothing but the cell adjacent to $\bi$ in direction $\vv{v}{}$, if
 $\vv{v}{}\in V_+$, or $\bi$ itself if $\vv{v}{}=\vv{0}{}$.

\subsection{The neighborhood}
\label{sect:neibor}

As mentioned in Section 1,  we only consider CAs with the von Neumann neighborhood. So, for each cell $\bi \in\C$, its neighborhood $P(\bi )$ consists of the cell $\bi$ and 
its $2d$ adjacent cells:
\[
P(\bi )= \bi + V := \{ \bi + \vv{v}{}\mid \vv{v}{}\in V \}\, .
\]

The results presented in \cite{NCCA} allow to describe the relation between $P(\bi )$ and $P(\bj )$ if $\bi \neq \bj$, as summarized in Lemma \ref{lemma:vN}.
\begin{lem}\label{lemma:vN}
Let $\bi ,\bj \in\C$ and $\bi \neq \bj $. 
\begin{itemize}
\item[(a)] If $\bj =\bi +\vv{v}{}$ for some $\vv{v}{}\in V_+$, then $P(\bi )\cap P(\bj )=\{ \bi+\vv{0}{} ,\bi+\vv{v}{}\}$.
\item[(b)] If $\bj = \bi +\vv{v}{} + \vv{v}{}$ for some $\vv{v}{}\in V_+$, then $P(\bi )\cap P(\bj )=\{ \bi+\vv{v}{} \}$. 
\item[(c)] If $\bj = \bi +\vv{u}{} + \vv{v}{}$
 for some $\vv{u}{},\vv{v}{}\in V_+$ where both $\vv{u}{}\neq \vv{v}{}$ and 
 $\vv{u}{}\neq \mbox{-}\vv{v}{}$, 
 then $P(\bi )\cap P(\bj)=\{\bi+\vv{u}{}, \bi +\vv{v}{}\}$.
\item[(d)] In all other cases, it holds that $P(\bi )\cap P(\bj )=\emptyset$.
\end{itemize}
\end{lem}

We now define a set $\PP$ containing all possible pairs of vectors from $V$ used in Lemma \ref{lemma:vN} in the description of $P(\bi )\cap P(\bj)$ for the cases (a) and (c). Thus $\PP$ consists of the pairs
 $\lbrace \vv{0}{}, \vv{v}{} \rbrace$
 for every $\vv{v}{}\in V_+$ and the pairs 
 $\lbrace \vv{u}{}, \vv{v}{}\rbrace$ for $\vv{u}{},\vv{v}{}\in V_+$ with $\vv{u}{}\neq \vv{v}{}$ and $\vv{u}{}\neq \mbox{-}\vv{v}{}$. 
 
As the pairs  $\lbrace \vv{u}{}, \vv{v}{}\rbrace$ and $\lbrace \vv{v}{}, \vv{u}{}\rbrace$ are equal, the set $\PP$ contains exactly $2d^2$ elements. 
For example, if $d=1$, then $\PP$ contains $2$ elements:
\begin{equation}
\PP = \bigg\lbrace  
\lbrace \vv{0}{},\vv{v}{1}\rbrace,   
\lbrace \vv{0}{},\mbox{-}\vv{v}{1}\rbrace   \bigg\rbrace,
\label{d1vectors}
\end{equation}
if $d=2$, then $\PP$ contains $8$ elements:
\begin{equation}
\PP = \bigg\lbrace  
\lbrace \vv{0}{},\vv{v}{1}\rbrace,  \lbrace\vv{0}{},\vv{v}{2}\rbrace ,  \lbrace \vv{v}{1},\vv{v}{2}\rbrace, 
\lbrace\vv{v}{1},\mbox{-}\vv{v}{2}\rbrace,
\lbrace \vv{0}{},\mbox{-}\vv{v}{1}\rbrace,  \lbrace\vv{0}{},\mbox{-}\vv{v}{2}\rbrace,  \lbrace\mbox{-}\vv{v}{1},\mbox{-}\vv{v}{2}\rbrace, \lbrace\mbox{-}\vv{v}{1},\vv{v}{2}\rbrace \bigg\rbrace\, ,
\label{d2vectors}
\end{equation}
while if $d=3$, then it holds that $|\PP|=18$:
\begin{equation}
\setlength{\arraycolsep}{0pt}
\begin{array}{l}
\PP = \bigg\lbrace
\lbrace\vv{0}{},\vv{v}{1}\rbrace,  \lbrace\vv{0}{},\vv{v}{2}\rbrace, 
 \lbrace\vv{0}{},\vv{v}{3}\rbrace,  \lbrace\vv{v}{1},\vv{v}{2}\rbrace,  \lbrace\vv{v}{1},\mbox{-}\vv{v}{2}\rbrace,  \lbrace\vv{v}{1},\vv{v}{3}\rbrace,  \lbrace\vv{v}{1},\mbox{-}\vv{v}{3}\rbrace,  \lbrace\vv{v}{2},\vv{v}{3}\rbrace,  \lbrace\vv{v}{2},\mbox{-}\vv{v}{3}\rbrace,\\
\lbrace\vv{0}{},\mbox{-}\vv{v}{1}\rbrace,  \lbrace\vv{0}{},\mbox{-}\vv{v}{2}\rbrace,  \lbrace\vv{0}{},\mbox{-}\vv{v}{3}\rbrace,  \lbrace\mbox{-}\vv{v}{1},\mbox{-}\vv{v}{2}\rbrace, \lbrace\mbox{-}\vv{v}{1},\vv{v}{2}\rbrace,  \lbrace\mbox{-}\vv{v}{1},\mbox{-}\vv{v}{3}\rbrace,  \lbrace\mbox{-}\vv{v}{1},\vv{v}{3}\rbrace,  \lbrace\mbox{-}\vv{v}{2},\mbox{-}\vv{v}{3}\rbrace,  \lbrace\mbox{-}\vv{v}{2},\vv{v}{3}\rbrace
\bigg\rbrace\,.
\end{array}
\label{d3vectors}
\end{equation}

From Lemma \ref{lemma:vN}, we also have the following.
\begin{rem}\label{remark}
Let $\bi\in\C$. If for some $\bj\in\C$ we have $|P(\bi ) \cap P(\bj )| =2$,
then there exists a unique element $\lbrace\vv{u}{}, \vv{v}{}\rbrace\in\PP$ such that 
$\bj =(\bi +\vv{u}{})+\vv{v}{}$. Moreover, one of the shared cells is $\bi +\vv{u}{} = \bj +(\mbox{-}\vv{v}{})$, while the other one is $\bi +\vv{v}{}
= \bj + (\mbox{-}\vv{u}{})$.
\end{rem}
For a given pair $\lbrace \vv{u}{}, \vv{v}{}\rbrace\in\PP$, the pair $\lbrace\mbox{-}\vv{u}{},
 \mbox{-}\vv{v}{}\rbrace$ will be called the \emph{matching pair}. For example, in the two-dimensional case,
$\lbrace\vv{0}{},\vv{v}{1}\rbrace$, $\lbrace\vv{0}{},\vv{v}{2}\rbrace$, $\lbrace\vv{v}{1},\vv{v}{2}\rbrace$, $\lbrace\vv{v}{1},\mbox{-}\vv{v}{2}\rbrace$ are the matching pairs of
$\lbrace \vv{0}{},\mbox{-}\vv{v}{1}\rbrace$,  $\lbrace\vv{0}{},\mbox{-}\vv{v}{2}\rbrace$, $\lbrace\mbox{-}\vv{v}{1},\mbox{-}\vv{v}{2}\rbrace$,  $\lbrace\mbox{-}\vv{v}{1},\vv{v}{2}\rbrace$, 
and vice versa.

It is easy to see that different elements from $\PP$ have different matching elements. 
If from each of the $d^2$ pairs of matching elements we choose one, then we get a set, which we denote by $\CP$. 
We can construct $\CP$ in $2^{d^2}$ ways, but it always holds that 
 $|\CP |=d^2$.

For the sake of convenience, in the examples presented in this paper, we choose $\CP = \bigg\lbrace  
\lbrace \vv{0}{},\vv{v}{1}\rbrace   \bigg\rbrace$,
if $d=1$, while if $d=2$, we choose 
\begin{equation}
\CP = \bigg\lbrace  
\lbrace \vv{0}{},\vv{v}{1}\rbrace,  \lbrace\vv{0}{},\vv{v}{2}\rbrace ,  \lbrace \vv{v}{1},\vv{v}{2}\rbrace, 
\lbrace\vv{v}{1},\mbox{-}\vv{v}{2}\rbrace \bigg\rbrace\, ,
\label{d2vectors-ch}
\end{equation}
and if $d=3$, we choose
\begin{equation}
\setlength{\arraycolsep}{0pt}
\begin{array}{l}
\CP=\bigg\lbrace
\lbrace\vv{0}{},\vv{v}{1}\rbrace, 
\lbrace\vv{v}{1},\mbox{-}\vv{v}{2}\rbrace, \lbrace\vv{v}{1},\vv{v}{3}\rbrace,  \lbrace\vv{v}{1},\mbox{-}\vv{v}{3}\rbrace,    
  \lbrace\vv{0}{},\mbox{-}\vv{v}{2}\rbrace,  \lbrace\vv{0}{},\mbox{-}\vv{v}{3}\rbrace,  \lbrace\mbox{-}\vv{v}{1},\mbox{-}\vv{v}{2}\rbrace,  \lbrace\mbox{-}\vv{v}{2},\mbox{-}\vv{v}{3}\rbrace,  \lbrace\mbox{-}\vv{v}{2},\vv{v}{3}\rbrace
\bigg\rbrace\,.
\end{array}
\label{CPd3vectors}
\end{equation}

\subsection{Configurations}

As the state set of a CA we consider a set $Q\subseteq\R$ (finite or not) containing zero and at least one more number. By $Q_+$ we denote the set $Q\setminus\{0\}$.
By a \emph{configuration}, we mean any mapping from the grid $\C$ to $Q$ and we denote the set of all configurations by $X=Q^{\C}$. To develop our idea of decomposition, it is important to consider \emph{configurations in a wider sense}: mappings from the grid $\C$ to $\R$. The set of all configurations in a wider sense is, of course, a superset of $X$ and is denoted by $\X$. 
We will simply write configuration also for elements from $\X$, unless confusion is possible.
The value of cell $\bi $ in a configuration $\bx\in \X$ is denoted by $\bx(\bi )$ or, if $d=2$, by $x_{i,j}$ for the cell $(i,j)$. 

Given a configuration $\bx\in \X$, we define its density as:
\[
\rho(\bx)=\frac{1}{|\C|}\sigma(\bx), \quad \text{where}\quad \sigma(\bx)=\sum_{\bi \in\C}\bx(\bi )\, .
\]
The set of all configurations $\bx\in \X$ satisfying $\sigma(\bx)=0$ is denoted by $\X_0$.

By a \emph{neighborhood configuration} we mean any function $N:\; V\to Q$. If $N$ is identically equal to zero, then we call it \emph{trivial}. The set of all possible neighborhood configurations is denoted by~$\NN$.
As the set $V$ has $2d+1$ elements, namely 
$\mbox{-}\vv{v}{d}$, $\ldots$, $\mbox{-}\vv{v}{1}$, $\vv{0}{}$, $\vv{v}{1}$, $\ldots$, $\vv{v}{d}$, we can define any neighborhood configuration by the sequence $N(\mbox{-}\vv{v}{d}), \ldots, N(\mbox{-}\vv{v}{1}), N(\vv{0}{}), N(\vv{v}{1}), \ldots, N(\vv{v}{d})$ (often without commas, unless confusion is possible). 
In the two-dimensional case, we prefer, however, to represent $N$ graphically as 
$
\setlength{\arraycolsep}{1pt}
\renewcommand{\arraystretch}{0.8}
\begin{array}{rcl}
 & q_1 &  \\
q_2 & q_3 & q_4, \\
 & q_5 & 
\end{array}$
with $N(\vv{0}{})=q_3$, $N(\vv{v}{1})=q_4$, $N(\mbox{-}\vv{v}{1})=q_2$, $N(\vv{v}{2})=q_1$ and $N(\mbox{-}\vv{v}{2})=q_5$. In this way,
the set $\NN$ can be defined as
\[
\NN=\left\{ 
\setlength{\arraycolsep}{1pt}
\renewcommand{\arraystretch}{0.8}
\begin{array}{rcl}
 & q_1 &  \\
q_2 & q_3 & q_4 \\
 & q_5 & 
\end{array}\mid q_1,q_2,q_3,q_4,q_5\in Q
\right\}.
\]
Similarly, for $d=3$, we write $\setlength{\arraycolsep}{1pt} 
\renewcommand{\arraystretch}{0.8}
\begin{array}{rcl}
 & q_1 & {\color{blue}q_7} \\
 q_2 & q_3 & q_4, \\
{\color{blue}q_6} & q_5 &  
\end{array}$ with $N(\vv{0}{})=q_3$, $N(\vv{v}{1})=q_4$, $N(\mbox{-}\vv{v}{1})=q_2$, $N(\vv{v}{2})=q_1$, $N(\mbox{-}\vv{v}{2})=q_5$, $N(\vv{v}{3})=q_6$ and $N(\mbox{-}\vv{v}{3})=q_7$. 

Some particular neighborhood configurations will be very important in remainder of this paper.
Firstly, for any $q\in Q$, we define the homogeneous neighborhood configuration $H_{q}$, in which for each $\vv{v}{}\in V$ we have $H_{q}(\vv{v}{})=q$.
Secondly, for any $\vv{v}{}\in V$ and $q\in Q$, we define a \emph{monomer} $M_{\vv{v}{}:q}$, which differs from $H_{0}$ in at most one component. More precisely, $M_{\vv{v}{}:q}$ has the value $q$ in direction $\vv{v}{}$ and zero in all other directions:
\[
\left( \forall \vv{u}{}\in V \right) \left(M_{\vv{v}{}:q}(\vv{u}{})=
\left\{\begin{array}{ll}
q, & \text{if }  \vv{u}{}= \vv{v}{}\\
0, & \text{if }  \vv{u}{}\neq \vv{v}{}
\end{array}\right.
\!\!\!\right)\, .
\]
Of course, if $q=0$, then $M_{\vv{v}{}:q}$ is trivial. 
Thus, if $d=1$, monomers are of the form $q00$, $0q0$ and $00q$, where $q\in Q$, while for $d=2$ and $d=3$ they look like:
\[
\setlength{\arraycolsep}{1pt}
\renewcommand{\arraystretch}{0.8}
\begin{array}{rcl}
 & q &  \\
0 & 0 & 0, \\
 & 0 & 
\end{array}\;
\setlength{\arraycolsep}{1pt}
\renewcommand{\arraystretch}{0.8}
\begin{array}{rcl}
 & 0 &  \\
q & 0 & 0, \\
 & 0 & 
\end{array}\;
\setlength{\arraycolsep}{1pt}
\renewcommand{\arraystretch}{0.8}
\begin{array}{rcl}
 & 0 &  \\
0 & q & 0, \\
 & 0 & 
\end{array}\;
\setlength{\arraycolsep}{1pt}
\renewcommand{\arraystretch}{0.8}
\begin{array}{rcl}
 & 0 &  \\
0 & 0 & q, \\
 & 0 & 
\end{array}\;  
\setlength{\arraycolsep}{1pt}
\renewcommand{\arraystretch}{0.8}
\begin{array}{rcl}
 & 0 &  \\
0 & 0 & 0 \\
 & q & 
\end{array}
\quad \text{and} \quad
\setlength{\arraycolsep}{1pt}
\renewcommand{\arraystretch}{0.8}
\begin{array}{rcl}
 & q & {\color{blue}0} \\
0 & 0 & 0, \\
{\color{blue}0} & 0 & 
\end{array}\;
\setlength{\arraycolsep}{1pt}
\renewcommand{\arraystretch}{0.8}
\begin{array}{rcl}
 & 0 & {\color{blue}0} \\
q & 0 & 0, \\
{\color{blue}0} & 0 & 
\end{array}\;
\setlength{\arraycolsep}{1pt}
\renewcommand{\arraystretch}{0.8}
\begin{array}{rcl}
 & 0 & {\color{blue}0} \\
0 & q & 0, \\
{\color{blue}0} & 0 & 
\end{array}\;
\setlength{\arraycolsep}{1pt}
\renewcommand{\arraystretch}{0.8}
\begin{array}{rcl}
 & 0 & {\color{blue}0} \\
0 & 0 & q, \\
{\color{blue}0} & 0 & 
\end{array}\;
\setlength{\arraycolsep}{1pt}
\renewcommand{\arraystretch}{0.8}
\begin{array}{rcl}
 & 0 & {\color{blue}0} \\
0 & 0 & 0, \\
{\color{blue}0} & q & 
\end{array}\;
\setlength{\arraycolsep}{1pt}
\renewcommand{\arraystretch}{0.8}
\begin{array}{rcl}
 & 0 & {\color{blue}0} \\
0 & 0 & 0, \\
{\color{blue}q} & 0 & 
\end{array}\;
\setlength{\arraycolsep}{1pt}
\renewcommand{\arraystretch}{0.8}
\begin{array}{rcl}
 & 0 & {\color{blue}q} \\
0 & 0 & 0, \\
{\color{blue}0} & 0 & 
\end{array}\;
\]
respectively.

Lastly, if $\lbrace \vv{u}{},\vv{w}{}\rbrace\in \PP$ and $p,q\in Q$, we define a  {\it dimer} $\dimer{\vv{u}{}:p}{\vv{w}{}:q}$ as the neighborhood configuration that has the value $p$ in direction $\vv{u}{}$, the value $q$ in direction $\vv{w}{}$ and zero in all other directions:
\[
\left(
\forall \vv{v}{}\in V \right) \left(\dimer{\vv{u}{}:p}{
\vv{w}{}:q}(\vv{v}{})=
\left\{\begin{array}{ll}
p, & \text{if }\vv{v}{}=\vv{u}{}\\
q, & \text{if }\vv{v}{}=\vv{w}{}\\
0, & \text{otherwise}
\end{array}\right.
\!\!\!\right)\, .
\]
Let us note that if $p=0$ or $q=0$, then $\dimer{\vv{u}{}:p}{\vv{w}{}:q}$ is a monomer. It is obvious that $\dimer{\vv{u}{}:p}{\vv{w}{}:q}$ equals $\dimer{\vv{w}{}:q}{\vv{u}{}:p}$.
As the pairs $\lbrace\vv{u}{}, \vv{w}{}\rbrace$ and $\lbrace \mbox{-}\vv{u}{}, \mbox{-}\vv{w}{}\rbrace$ 
are called matching (see Section \ref{sect:neibor}), we also use this term to refer to the dimers $\dimer{\vv{u}{}:p}{\vv{w}{}:q}$ and $\dimer{\mbox{-}\vv{w}{}:p}{\mbox{-}\vv{u}{}:q}$. 
If $d=1$, dimers are of the form: $pq0$ and $0pq$, where $p, q\in Q$, while for $d=2$ they look like: 
\[
\setlength{\arraycolsep}{1pt}
\renewcommand{\arraystretch}{0.8}
\begin{array}{rcl}
 & 0 &  \\
0 & p & q, \\
 & 0 & 
\end{array}\;
\setlength{\arraycolsep}{1pt}
\renewcommand{\arraystretch}{0.8}
\begin{array}{rcl}
 & 0 &  \\
p & q & 0, \\
 & 0 & 
\end{array}\;
\setlength{\arraycolsep}{1pt}
\renewcommand{\arraystretch}{0.8}
\begin{array}{rcl}
 & p &  \\
0 & q & 0, \\
 & 0 & 
\end{array}\;
\setlength{\arraycolsep}{1pt}
\renewcommand{\arraystretch}{0.8}
\begin{array}{rcl}
 & 0 &  \\
0 & p & 0, \\
 & q & 
\end{array}\;
\setlength{\arraycolsep}{1pt}
\renewcommand{\arraystretch}{0.8}
\begin{array}{rcl}
 & p &  \\
q & 0 & 0, \\
 & 0 & 
\end{array}\;
\setlength{\arraycolsep}{1pt}
\renewcommand{\arraystretch}{0.8}
\begin{array}{rcl}
 & 0 &  \\
0 & 0 & p, \\
 & q & 
\end{array}\;
\setlength{\arraycolsep}{1pt}
\renewcommand{\arraystretch}{0.8}
\begin{array}{rcl}
 & p &  \\
0 & 0 & q, \\
 & 0 & 
\end{array}\;
\setlength{\arraycolsep}{1pt}
\renewcommand{\arraystretch}{0.8}
\begin{array}{rcl}
 & 0 &  \\
p & 0 & 0. \\
 & q & 
\end{array}
\]
Note that the states $p$ and $q$ occur at specific locations as given by the set $\PP$. 

\noindent
If $\bx\in X$ and $\bi \in\C$ are given, then $N_{\bx,\bi}$ denotes the neighborhood configuration given by the von Neumann neighborhood of cell $\bi$ in the configuration $\bx$, thus
\[
\left(\forall \vv{v}{}\in V \right) \left( N_{\bx,\bi} (\vv{v}{})=\bx(\bi +\vv{v}{})\right).
\]

\subsection{Local and global functions and rules}

Any function $f:\; \NN\to \R$ is called a~\emph{local function}. If additionally $f(\NN)\subseteq Q$, then we call $f$ a \emph{local rule}. 
As for configurations, we generalize local rules in the sense that they may take values outside the state set $Q$.

Each local function $f$ induces a \emph{global function} $A_f: X\to \X$ defined for $\bx\in X$ and $\bi\in\C$ as follows
\[
A_f(\bx)(\bi )=f(N_{\bx,\bi }).
\]
If $f$ is a local rule, then we call $A_f$ a global rule and then $A_f: X\to X$.

Having introduced the required notations, we now define the property of being \emph{number-conserving}.
\begin{defn}
A local function $f$ is called number-conserving if its corresponding global function $A_f$ conserves density, \emph{i.e.}, for each $\bx \in X$ it holds that $\rho(A_f(\bx))=\rho(\bx)$, or, equivalently, that
$\sigma(A_f(\bx))=\sigma(\bx)$.  
\end{defn}
Note that if the state set contains not only natural numbers, then the term  \emph{density-conserving} is preferred. 

The following facts follow immediately.
\begin{lem}\label{lemma:qq3}
If a local function $f$ is number-conserving, then each state is quiescent, \emph{i.e.}, for any state $q\in Q$ it holds that $f(H_q)=q$. 
\end{lem}

\begin{lem}\label{lemma:qqq3}
If a local function $f$ is number-conserving, then for any state $q\in Q$ it holds that 
\[
\sum_{\vv{v}{}\in V}f(M_{\vv{v}{}:q}) = q.
\]
\end{lem}
The proofs of both Lemmata \ref{lemma:qq3} and \ref{lemma:qqq3} are exactly the same as those presented in \cite{NCCA} for local rules, so they are omitted.

\section{The decomposition of a number-conserving local rule}
\label{sec:decom}

We first present two fundamental objects underlying the results obtained in the remainder of this paper. The first is a \emph{split function} -- a special local function that is defined by the values it takes on monomers, and the second is  a \emph{perturbation} -- a local function that is trivial on monomers and whose induced global function maps every configuration to $\X_0$. It turns out that any number-conserving local rule can be decomposed into two such objects and that this decomposition is unique. 
 
\subsection{Split functions}

\begin{figure}[!ht]
\centering
\includegraphics[width=0.9\textwidth]{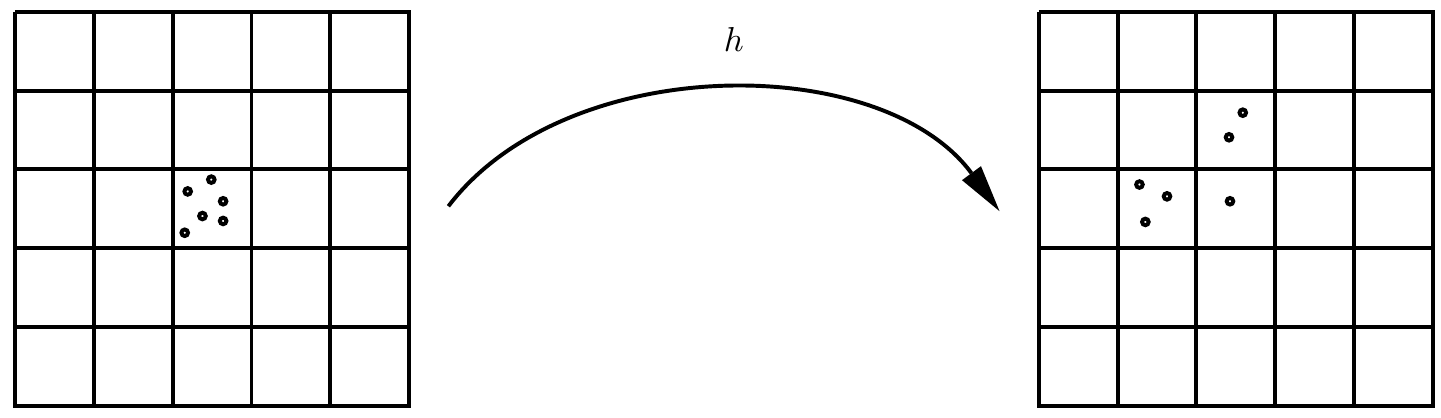}
\caption{The state $6$ in the central cell of a $5\times 5$ grid (symbolizing six particles) splits according to the following recipe: $3$ to the left neighbor, $2$ to the upper neighbor and $1$ stays in the cell. In the language of CAs, this means that $h(M_{\vv{v}{1}:6})=3$, $h(M_{\mbox{-}\vv{v}{2}:6})=2$, $h(M_{\vv{0}{}:6})=1$ and $h(M_{\mbox{-}\vv{v}{1}:6})=h(M_{\vv{v}{2}:6})=0$.}
\label{fig:out-in}
\end{figure}

Let us consider a configuration in which only one cell has some nonzero state and the other ones have state $0$. If a local rule would be number-conserving, then in the next time step it must redistribute this state to the cells located in its neighborhood in such a way that the redistributed parts belong to the state set $Q$. We say that this state splits, which happens according to some recipe depending on the state. Now, we are interested only in such local functions, called \emph{split functions}, that act as follows: each state splits according to its recipe irrespective of the states of its neighbors. We include these ideas in the definition below, reformulated in the language of CAs (see Figure~\ref{fig:out-in}).

\begin{defn}\label{split}
A local function $h$ is called  a split function if it satisfies:
\begin{itemize}
\item[{\rm (S1)}] $h(M_{\vv{v}{}:q})\in Q$, for any monomer $M_{\vv{v}{}:q}$;
\item[{\rm (S2)}] for any $q\in Q$, it holds that $\displaystyle{\sum_{\vv{v}{}\in V}h(M_{\vv{v}{}:q}) = q}$;
\item[{\rm (S3)}] for any $N\in \NN$, it holds that $\displaystyle{h(N) = \sum_{\vv{v}{}\in V}h(M_{\vv{v}{}:N(\vv{v}{})})}$.
\end{itemize} 
The set of all split functions is denoted by $\SF$.
\end{defn}
A split function is unambiguously defined by the values it takes on monomers, because its value on any neighborhood configuration is given by property (S3). Moreover, on monomers it takes values from the state set as a consequence of property (S1). However, a split function does not have to be a local rule. Indeed, as each state splits independently, it may happen that the sum of the splitted constituents ending up in one cell from different neighbors does not belong to the state set $Q$. So, $A_h:X\to \X$. Yet, we do not care about this for the time being.

The next lemma shows that each split function is number-conserving.
\begin{lem}\label{lemma:Mfunction}
Let $h$ be a split function. Then $\sigma (A_h(\bx)) = \sigma (\bx)$ for any configuration $\bx\in X$.
\end{lem}
Proof: Let a local function $h$ belong to $\SF$. According to property (S3), we get
\[
\sigma (A_h(\bx)) = \sum_{\bi\in\C}A_h(\bx)(\bi) =
\sum_{\bi\in\C} h(N_{\bx,\bi})=
\sum_{\bi\in\C}\sum_{\vv{v}{}\in V}h(M_{\vv{v}{}:N_{\bx,\bi}(\vv{v}{})}) =
\sum_{\bi\in\C}\sum_{\vv{v}{}\in V}h(M_{\vv{v}{}:\bx(\bi+\vv{v}{})})\, . 
\]
Next, using the fact that $\C + \vv{v}{}=\C$ for any $\vv{v}{}\in V$, we have
\[
\sum_{\bi\in\C}\sum_{\vv{v}{}\in V}h(M_{\vv{v}{}:\bx(\bi+\vv{v}{})})
= \sum_{\vv{v}{}\in V}\sum_{\bi\in\C} h(M_{\vv{v}{}:\bx(\bi+\vv{v}{})}) 
= \sum_{\vv{v}{}\in V}\sum_{\bi\in\C} h(M_{\vv{v}{}:\bx(\bi)}) 
= \sum_{\bi\in\C}\sum_{\vv{v}{}\in V} h(M_{\vv{v}{}:\bx(\bi)})\, .
\]
Finally, from property (S2), we get
$
\sum_{\bi\in\C}\sum_{\vv{v}{}\in V} h(M_{\vv{v}{}:\bx(\bi)}) = \sum_{\bi\in\C} \bx(\bi)
= \sigma (\bx),
$
which concludes the proof.
\hfill\ensuremath{\Box}

Note that the set $\SF$ has a very simple structure. Indeed, each split function is determined by the values it takes on monomers $M_{\vv{v}{}:q}$, where $\vv{v}{}\in V$ and $q\in Q_+$, so we can identify a split function $h$  with  
\[
\Big(\big( h(M_{\mbox{-}\vv{v}{d}:q}),\ldots ,h(M_{\mbox{-}\vv{v}{1}:q}),h(M_{\vv{0}{}:q}),h(M_{\vv{v}{1}:q}),\ldots ,h(M_{\vv{v}{d}:q})\big)\Big)_{q\in Q_+}\, .
\]
According to properties (S1) and (S2), each tuple
\begin{equation}\label{tuple}
\big( h(M_{\mbox{-}\vv{v}{d}:q}),\ldots ,h(M_{\mbox{-}\vv{v}{1}:q}),h(M_{\vv{0}{}:q}),h(M_{\vv{v}{1}:q}),\ldots ,h(M_{\vv{v}{d}:q})\big) 
\end{equation}
belongs to the set $S_q,$ defined as
\[
S_q=\big\{ (x_1,x_2,\ldots ,x_{2d+1}) \mid x_1+x_2+\ldots +x_{2d+1}=q,\; x_1,x_2,\ldots ,x_{2d+1}\in Q\big\}\, .
\]
Since for each state $q$ the corresponding tuple in (\ref{tuple}) can be chosen independently, the set of all split functions can be identified with the Cartesian product $\displaystyle{\bigtimes_{q\in Q_+}S_q}$. In particular, if the state set $Q$ is finite, the cardinality of the set $\SF$ equals $\displaystyle{\prod_{q\in Q_+}|S_q |}$.

To better understand the meaning of a split function, we consider a classical CAs where the state set is a discrete set $\{0,1,\ldots,\qm\}$ for some natural number $\qm$. One can see that the set $\SF$ is very simple to describe and to find. 

\begin{example}\label{example:q}
Multi-state CAs
\end{example}
Let us assume that $Q=\{0,1,\ldots,q^*\}$ for some positive integer $q^*$ and let the dimension $d$ be given.
The cardinality of the set of all local rules in this case equals $(q^*+1)^{(q^*+1)^{2d+1}}$, so even for very small $q^*$ and $d$ it can be huge (see Table~\ref{tab:SF}(a)).
However, the cardinality of $\SF$ is equal to $k_1\cdot k_2 \cdot\ldots\cdot k_{q^*}$, where for $q\in \{1,\ldots,q^*\}$ the symbol $k_q$ denotes the number of solutions of the equation $x_1+x_2+\ldots +x_{2d+1}=q$ in the set $Q$. 
Indeed, as we mentioned above, according to properties (S1) and (S2), one gets all split functions in this case by finding for each $q\in Q$ all possibilities for $h(M_{\mbox{-}\vv{v}{d}:q})$, $\ldots$, $h(M_{\mbox{-}\vv{v}{1}:q})$,  $h(M_{\vv{0}{}:q})$, $h(M_{\vv{v}{1}:q})$, $\ldots$, $h(M_{\vv{v}{d}:q}) \in Q$ such that
\[
h(M_{\mbox{-}\vv{v}{d}:q}) + \ldots + h(M_{\mbox{-}\vv{v}{1}:q}) +   h(M_{\vv{0}{}:q}) + h(M_{\vv{v}{1}:q}) + \ldots + h(M_{\vv{v}{d}:q}) = q\, .
\]
As $k_q = {\binom{2d+q}{q}}$ (see e.g.~\cite{ross2011first}), we have
\begin{equation}
\label{eq:S}
|\SF |= { \binom{2d+1}{1} } { \binom{2d+2}{2} }\cdot\ldots\cdot { \binom{2d+q^*}{q^*} }.
\end{equation}
Table~\ref{tab:SF}(b) presents $|\SF|$ for $d\leq 4$ and $q^*\leq 3$. One can see that the total number of split functions is definitely less than the number of all local rules. For example, in the two-dimensional case for three states (\emph{i.e.}, $q^*=2$), there are only $75$ split functions, while there are $3^{3^5}$ local rules  -- the last number has $116$ digits.

\begin{table}[!htbp]
\begin{center}
\subfloat[]{
\begin{tabular}{|c|c|c|c|c|c|}
		\hline
		 \rule{0pt}{1.2em} & $d=1$ & $d=2$ & $d=3$ & $d=4$  \\
		\hline
	     \rule{0pt}{1.2em}$q^*=1$ & 256 & $2^{32}$ & $2^{128}$ & $2^{512}$ \\
		\hline
	     \rule{0pt}{1.2em}$q^*=2$ & $3^{27}$ & $3^{243}$ & $3^{2187}$ & $3^{19683}$ \\
	\hline
	     \rule{0pt}{1.2em}$q^*=3$ & $4^{64}$ & $4^{1024}$ & $4^{16384}$ & $4^{262144}$ \\
		\hline
	\end{tabular}
	}
	\subfloat[]{
	\begin{tabular}{|c|c|c|c|c|c|}
		\hline
		 \rule{0pt}{1.2em} & $d=1$ & $d=2$ & $d=3$ & $d=4$  \\
		\hline
	     \rule{0pt}{1.2em}$q^*=1$ & 3 & 5 & 7 & 9 \\
		\hline
	     \rule{0pt}{1.2em}$q^*=2$ & 18 & 75 & 196 & 405 \\
	\hline
	     \rule{0pt}{1.2em}$q^*=3$ & 180 & 2625 & 16464 & 66825 \\
		\hline
	\end{tabular}
	}
	\caption{The number of all local rules (a) versus the number of all split functions (b) for state set $Q=\{0,1,\ldots,q^*\}$ and dimension $d$.}
    \label{tab:SF}
\end{center}
\end{table}

\subsection{Perturbations}

Now, we introduce the definition of the second important kind of local function, a so-called \emph{perturbation}. As we will see later, for a number-conserving local rule, a perturbation is the only possible derogation from being a split function. We will show that the set of all perturbations has a very nice structure -- it is a linear space -- and therefore it can be very easily described in terms of the elements of a basis of this space.

\begin{defn}\label{perturbation}
A local function $g:\NN\to\R$ is called a \emph{perturbation} if it satisfies the following two conditions:
\begin{itemize}
\item[{\rm (P1)}] for any $\vv{v}{}\in V$ and for any $q\in Q$, it holds that $\displaystyle{g(M_{\vv{v}{}:q}) = 0}$,
\item[{\rm (P2)}] for any $\bx\in X$, it holds that $\displaystyle{\sigma(A_g(\bx))=0}$.
\end{itemize}  
The set of all perturbations is denoted by $\Per$.
\end{defn}
Note that condition (P1) ensures that perturbations always take values zero on monomers, while condition (P2) means that $g: X \to \X_0$, for any $g\in\Per$. 
\begin{lem}\label{lemma:pert1}
The set of all perturbations $\Per$ forms a linear space.
\end{lem}
Proof: Let $g_1, g_2\in\Per$ and let $\alpha$, $\beta$ be arbitrary real numbers. Then for $g=\alpha g_1 + \beta g_2$, we have:
\[
g(M_{\vv{v}{}:q}) = \alpha g_1(M_{\vv{v}{}:q}) + \beta g_2(M_{\vv{v}{}:q}) = 0+0=0,
\]
for any $\vv{v}{}\in V$ and $q\in Q$. Moreover, for any $\bx\in X$ we have
\[
\sigma(A_g(\bx)) = \sum_{\bi\in\C}A_g(\bx)(\bi) =
\sum_{\bi\in\C} g(N_{\bx,\bi})= 
\sum_{\bi\in\C} \left(\alpha g_1(N_{\bx,\bi})+ \beta g_2(N_{\bx,\bi})\right)
\]
\[
= \alpha \sum_{\bi\in\C} g_1(N_{\bx,\bi}) + \beta \sum_{\bi\in\C} g_2(N_{\bx,\bi})
=
\alpha\sum_{\bi\in\C}A_{g_1}(\bx)(\bi) + \beta\sum_{\bi\in\C}A_{g_2}(\bx)(\bi) = 0 +0 = 0,
\]
which implies that $g$ is a perturbation.
\hfill\ensuremath{\Box}

The next lemma shows the relationship between the values of a perturbation on matching dimers. It will be used to characterize all perturbations and to find a basis of $\Per$. 
\begin{lem}\label{lemma:pert2}
If a local function $g:\NN\to\R$ is a perturbation, then for any $
\lbrace \vv{u}{},\vv{w}{}\rbrace \in \PP$ and for any $p,q\in Q$ it holds that 
\[
g\left( \dimer{\vv{u}{}:p}{
\vv{w}{}:q}\right) = - g\left( \dimer{\mbox{-}\vv{w}{}:p}{
\mbox{-}\vv{u}{}:q}\right).
\]
\end{lem}
Proof: Let $
\lbrace \vv{u}{},\vv{w}{}\rbrace \in \PP$ and $p,q\in Q$ be given. Let
us define the following configuration $\bx\in X$:
\[
\bx (\bi )=
\begin{cases}
p,  \quad\text{if } \bi = \bze+\vv{u}{}\\
q, \quad\text{if } \bi = \bze+\vv{w}{}\\
0,  \quad\text{otherwise.}\\
\end{cases}
\]
We have that
\[
\sigma(A_g(\bx)) = \sum_{\bi\in\C}A_g(\bx)(\bi) =
\sum_{\bi\in\C} g(N_{\bx,\bi})\, . 
\]
If $\bze+\vv{u}{}$ or $\bze+\vv{w}{}$ does not belong to the neighborhood $P(\bi)$, then $N_{\bx,\bi }$ is a monomer, so in this case $g(N_{\bx,\bi})=0$.
On the other hand, $P(\bi)$ contains $\bze+\vv{u}{}$ if and only if $\bze+\vv{u}{} =\bi +\vv{v}{}$ for some $\vv{v}{}\in V$, and, more importantly, we then have $N_{\bx,\bi}(\vv{v}{})=N_{\bx,\bze + \vv{u}{} - \vv{v}{} }(\vv{v}{})=\bx (\bze+\vv{u}{})$. 
Similarly, $P(\bi)$ contains $\bze+\vv{w}{}$ if and only if $\bze+\vv{w}{} =\bi +\vv{v}{}'$ for some $\vv{v}{}'\in V$ and then $N_{\bx,\bi } (\vv{v}{}')=N_{\bx,\bze+\vv{w}{}-\vv{v}{}'}(\vv{v}{}')=\bx (\bze+\vv{w}{})$. 
Thus $P(\bi )$ contains both $\bze+\vv{u}{}$ and $\bze+\vv{w}{}$ only when  $\bi = \bze + \vv{u}{} - \vv{v}{}=
 \bze + \vv{w}{} - \vv{v}{}'$ for some $\vv{v}{},\vv{v}{}'\in V$. 
As $\lbrace \vv{u}{},\vv{w}{}\rbrace \in \PP$, $\vv{u}{}$ and $\vv{w}{}$ 
act on different components and at least one of them is nonzero. Hence, the vector equation 
$\bi = \bze + \vv{u}{} - \vv{v}{}= \bze + \vv{w}{} - \vv{v}{}'$
 has exactly two solutions:
\[
\left\{\begin{array}{l}
\vv{v}{} =\mbox{-}\vv{w}{}\\
\vv{v}{}'= \mbox{-}\vv{u}{}
\end{array}\right.
\quad{} \text{ or } \quad{}
\left\{\begin{array}{l}
\vv{v}{}=\vv{u}{}\\
\vv{v}{}'=\vv{w}{}.
\end{array}\right.
\]
From this, we conclude that $P(\bi )$ contains both $\bze+\vv{u}{}$ and $\bze+\vv{w}{}$ only when $\bi=\bze$ or 
$\bi=\bze+\vv{u}{}+\vv{w}{}$. 
Moreover, we have
\begin{equation}\label{dimers}
N_{\bx,\bze }=\dimer{\vv{u}{}:\bx (\bze+\vv{u}{})}{\vv{w}{}:\bx (\bze+\vv{w}{})} =\dimer{\vv{u}{}:p}{
\vv{w}{}:q} \quad\text{ and }\quad N_{\bx,\bze+\vv{u}{}+\vv{w}{} }=
\dimer{\mbox{-}\vv{w}{}:\bx (\bze+\vv{u}{})}{\mbox{-}\vv{u}{}:\bx (\bze+\vv{w}{})}=\dimer{\mbox{-}\vv{w}{}:p}{
\mbox{-}\vv{u}{}:q}\, .
\end{equation}
Summarizing the above observations, we get
\[
\sigma(A_g(\bx)) =  g(N_{\bx,\bze}) + g(N_{\bx,\bze+\vv{u}{}+\vv{w}{}}) = 
g\left(\dimer{\vv{u}{}:p}{
\vv{w}{}:q}\right) + g\left(\dimer{\mbox{-}\vv{w}{}:p}{
\mbox{-}\vv{u}{}:q}\right)\, .
\]
Since property (P2) guarantees that $\sigma(A_g(\bx))=0$, we get our claim. 
\hfill\ensuremath{\Box}

Next we obtain a necessary and sufficient condition for a local function to be a perturbation. Its notation depends on our choice of $\CP$, so, there are $2^{d^2}$ equivalent formulations of it (see also Section~2.2).

\begin{thm}\label{main-pert}
Let $\CP$ be fixed. A local function $g$ is a perturbation if and only if it satisfies {\rm (P1)} of Definition \ref{perturbation} and for any
$N\in\NN$ it holds that
\begin{equation}
\label{eq:thm-pert}
g(N) = \sum_{\lbrace\vv{u}{},\vv{w}{}\rbrace\in \CP}\left[ 
g\left( \dimer{\vv{u}{}:N(\vv{u}{})}
{\vv{w}{}:N(\vv{w}{})}\right)
 -
g\left(  \dimer{\vv{u}{}:N(\mbox{-}\vv{w}{})}
{\vv{w}{}:N(\mbox{-}\vv{u}{})}\right)\right] .
\end{equation}
\end{thm}
Proof: 
It is easy to see that this condition is sufficient. Indeed, consider the grid $\C$ and an arbitrary configuration $\bx\in X$. Now,  write \eqref{eq:thm-pert} for each neighborhood configuration $N_{\bx,\bi}$. Summing up the left-hand sides, we obtain  $\sigma (A_g(\bx ))$, while when summing up the right-hand sides, the dimers cancel out due to the fact that  $N_{\bx,\bi+\vv{u}{}+\vv{w}{}}(\mbox{-}\vv{w}{})=\bx (\bi + \vv{u}{})=N_{\bx,\bi}(\vv{u}{})$ and $N_{\bx,\bi+\vv{u}{}+\vv{w}{}}(\mbox{-}\vv{u}{})=\bx (\bi + \vv{w}{})=N_{\bx,\bi}(\vv{w}{})$, so 
\[ \dimer{\vv{u}{}:N_{\bx,\bi}(\vv{u}{})}
{\vv{w}{}:N_{\bx,\bi}(\vv{w}{})} = \dimer{\vv{u}{}:N_{\bx,\bj}(\mbox{-}\vv{w}{})}
{\vv{w}{}:N_{\bx,\bj}(\mbox{-}\vv{u}{})}\, ,
\]
when $\bj=\bi+\vv{u}{}+\vv{w}{}$.

To prove that Eq.~(\ref{eq:thm-pert}) is necessary, let us fix $N\in \NN$ and consider the configuration $\bx$ defined by
\begin{equation*}
\bx (\bi )=
\left\{\begin{array}{ll}
N(\vv{v}{}), & \text{if }\ \bi = \bze +\vv{v}{} \text{ for some } \vv{v}{}\in V\\
0,  & \text{otherwise.}
\end{array}\right.
\end{equation*}
Note that $N_{\bx ,\bze }=N$. As a configuration $\bx$ is zero outside $P(\bze)$, from Lemma \ref{lemma:vN} we have
\begin{equation}
\label{eqSigmaF}
0 = \sigma (A_g(\bx ))=\sum_{\bi\in\C} g(N_{\bx,\bi })=g(N_{\bx,\bze})+
\sum_{\{\bi \mid\; |P(\bi )\cap P(\bze )|=2\}}\hspace{-6mm} g(N_{\bx,\bi })+\sum_{\{\bi \mid\; |P(\bi )\cap P(\bze )|=1\}}\hspace{-6mm} g(N_{\bx,\bi }).
\end{equation}
If $|P(\bi )\cap P(\bze )|=1$, then $N_{\bx,\bi }$ is a monomer, thus according to property (P1), we have that $g(N_{\bx,\bi })=0$. Moreover, in view of Remark \ref{remark}, if $|P(\bi )\cap P(\bze )|=2$, then there exists a unique pair $\lbrace\vv{u}{},\vv{w}{}\rbrace\in \PP$ such that $\bi =\bze + \vv{u}{} + \vv{w}{}$ and $P(\bi )\cap P(\bze )=\{\bze + \vv{u}{},\bze + \vv{w}{}\}$. Hence, $N_{\bx,\bi }$ is a dimer satisfying
\begin{equation}
\label{dimerxi}
N_{\bx,\bze + \vv{u}{} + \vv{w}{}}(\mbox{-}\vv{u}{})=\bx (\bze + \vv{w}{})=N(\vv{w}{})
\quad \text{ and }\quad
N_{\bx,\bze + \vv{u}{} + \vv{w}{}}(\mbox{-}\vv{w}{})=\bx (\bze + \vv{u}{})=N(\vv{u}{}),
\end{equation}
which means that $N_{\bx,\bze + \vv{u}{} + \vv{w}{}}=
\dimer{\mbox{-}\vv{w}{}:N(\vv{u}{})}{\mbox{-}\vv{u}{}:N(\vv{w}{})}$. Hence, for a given $\CP$, we have
\begin{equation}
\label{eqFromRk22}
\sum_{\{\bi \mid\; |P(\bi )\cap P(\bze )|=2\}} \hspace{-6mm} g(N_{\bx,\bi })=\hspace{-3mm} \sum_{\lbrace\vv{u}{},\vv{w}{}\rbrace\in \PP} \hspace{-1mm} g\left( \dimer{\mbox{-}\vv{w}{}:N(\vv{u}{})}{\mbox{-}\vv{u}{}:N(\vv{w}{})}
\right)=
\sum_{\lbrace\vv{u}{},\vv{w}{}\rbrace\in \CP}\left[  g\left( \dimer
{\mbox{-}\vv{w}{}:N(\vv{u}{})}{\mbox{-}\vv{u}{}:N(\vv{w}{})}\right) + g\left( \dimer
{\vv{u}{}:N(\mbox{-}\vv{w}{})}{\vv{w}{}:N(\mbox{-}\vv{u}{})}\right)\right] ,
\end{equation}
in agreement with the relation between $\PP$ and $\CP$ (see Section \ref{sect:neibor}).
Combining Eqs. (\ref{eqSigmaF}) and (\ref{eqFromRk22}) and recalling that $N_{\bx,\bze }=N$, we obtain:
\[
g(N) = - \sum_{\lbrace\vv{u}{},\vv{w}{}\rbrace\in \CP}\left[  g\left( \dimer
{\mbox{-}\vv{w}{}:N(\vv{u}{})}{\mbox{-}\vv{u}{}:N(\vv{w}{})}\right) + g\left( \dimer
{\vv{u}{}:N(\mbox{-}\vv{w}{})}{\vv{w}{}:N(\mbox{-}\vv{u}{})}\right)\right] .
\]

Since from Lemma \ref{lemma:pert2}, we know that
$ g\left( \dimer
{\mbox{-}\vv{w}{}:N(\vv{u}{})}{\mbox{-}\vv{u}{}:N(\vv{w}{})}\right) = -
g\left( \dimer
{\vv{u}{}:N(\vv{u}{})}{\vv{w}{}:N(\vv{w}{})}\right)$, the theorem is proven. 
\hfill\ensuremath{\Box}

Let us recall that if $p=0$ or $q=0$ then $\dimer{\vv{u}{}:p}{\vv{w}{}:q}$ is a monomer. As a consequence of Theorem \ref{main-pert} we obtain the following remark.
\begin{rem}
To define a perturbation, it suffices to declare its value on all dimers $\dimer
{\vv{u}{}:p}{\vv{w}{}:q}$, where 
$\lbrace\vv{u}{},\vv{w}{}\rbrace\in \CP$ and $p,q\in Q_+$.
\end{rem}

To illustrate the concept of a perturbation, we continue with the multi-state CAs presented in Example \ref{example:q}. We will see that the basis of $\Per$ is easy to find and describe.

\begin{example}\label{example:q-cd}
Multi-state CAs -- cont.
\end{example}
Since $|\CP|=d^2$ and $Q_+=\{1,\ldots,q^*\}$, the cardinality of the set 
\[\D = \left\{ \dimer
{\vv{u}{}:p}{\vv{w}{}:q} \mid \lbrace\vv{u}{},\vv{w}{}\rbrace\in \CP,\; p,q\in Q_+ \right\}
\]
is $d^2 (q^*)^2$.
Let $g_{\vv{u}{}:p,\vv{w}{}:q}$ denote the perturbation that maps to $1$ on the dimer $\dimer{\vv{u}{}:p}{\vv{w}{}:q}$ and zero on the other dimers in $\D$ and recall that it is zero on monomers too, while on other $N\in\NN$ it is given by Eq.~(\ref{eq:thm-pert}). Then the set
\[
\left\{ 
g_{\vv{u}{}:p,\vv{w}{}:q} \mid \lbrace\vv{u}{},\vv{w}{}\rbrace\in \CP,\; p,q\in Q_+
\right\}
\]
is a basis of $\Per$, so the dimension of $\Per$ equals $d^2 (q^*)^2$.
 In Table~\ref{tab:Per} the dimension of $\Per$ is given for $d\leq 4$ and $q^*\leq 3$. 

\begin{table}[ht]
	\centering
	\begin{tabular}{|c|c|c|c|c|c|}
		\hline
		 \rule{0pt}{1.2em} & $d=1$ & $d=2$ & $d=3$ & $d=4$  \\
		\hline
	     \rule{0pt}{1.2em}$q^*=1$ & 1 & 4 & 9 & 16 \\
		\hline
	     \rule{0pt}{1.2em}$q^*=2$ & 4 & 16 & 36 & 64 \\
	\hline
	     \rule{0pt}{1.2em}$q^*=3$ & 9 & 36 & 81 & 144 \\
		\hline
	\end{tabular}
		\caption{The dimension of the linear space $\Per$ for $Q=\{0,1,\ldots,q^*\}$ and dimension $d$.}
	\label{tab:Per}
\end{table}

\subsection{The decomposition theorem}

Now, we are ready to show that each number-conserving local rule can be decomposed as the sum of a split function and a perturbation.

\begin{thm}
\label{main-dec}
A local rule $f$ is number-conserving if and only if there exist a split function $h$ and a perturbation $g$ such that
$f=h+g$.
Moreover, for a given local rule $f$, the functions $h$ and $g$ are uniquely determined.
\end{thm}
Proof: 
If a local rule $f$ is the sum of some split function $h$ and a perturbation $g$, then it conserves density. Indeed, let $\bx\in X$ be any configuration, then
\[
\sigma(A_f(\bx)) = \sum_{\bi\in\C}A_f(\bx)(\bi) = \sum_{\bi\in\C}\left( A_h(\bx)(\bi) + A_g(\bx)(\bi)\right) = \sum_{\bi\in\C}A_h(\bx)(\bi) + \sum_{\bi\in\C}A_g(\bx)(\bi) = \sigma(\bx), 
\]
as each split function conserves density (Lemma \ref{lemma:Mfunction}) and $\displaystyle{\sum_{\bi\in\C}A_g(\bx)(\bi)=0}$, for any perturbation $g$ (Definition \ref{perturbation}).

Now, let us assume that a local rule $f$ is number-conserving. Let us define the local function $h$ as follows:
\begin{itemize}
\item[(i)] for any $\vv{v}{}\in V$ and for any $q\in Q$, we set $\displaystyle{h(M_{\vv{v}{}:q}) = f(M_{\vv{v}{}:q})}$,
\item[(ii)] for any $N\in \N$, we define $\displaystyle{h(N) = \sum_{\vv{v}{}\in V}h(M_{\vv{v}{}:N(\vv{v}{})})}$.
\end{itemize} 
It is easy to see that $h$ is a split function. Let $g=f-h$. As both $f$ and $h$ conserve density, for any $\bx\in X$, we have that $\sigma(A_g(\bx)) = 0$. Moreover, given the definition of $h$, for any $\vv{v}{}\in V$ and for any $q\in Q$, it holds that $g(M_{\vv{v}{}:q}) = 0$. So, $g$ is a perturbation.

The decomposition of a number-conserving local rule into a split function and a perturbation is unique. Indeed, if for split functions $h_1$ and $h_2$ and perturbations $g_1$ and $g_2$, we have $h_1 +g_1 = h_2 + g_2$, then $h_1  = h_2 $ on monomers, and therefore on the entire $\NN$, because split functions are defined by their values on monomers. Consequently, it must hold that also $g_1  = g_2$.
\hfill\ensuremath{\Box}

\section{Use cases}\label{sec:uses}
To  convince the reader that the decomposition theorem is useful, we show how it allows to reduce the computational complexity of finding all number-conserving CAs. For this purpose, we list all number-conserving rules for classical examples of $d$ and $Q$, which until now could only be done using a computer. Additionally, we rely on the decomposition theorem to find all three-dimensional three-state number-conserving CAs, which until now was beyond the capabilities of computers.

\subsection{Number-conserving binary CAs}
\label{subsec:NCCA}

{\it One-dimensional binary CAs} \\
Here we consider the simplest CAs, namely Elementary Cellular Automata (ECAs), which have been intensively investigated. In particular, it is known which ones are number-conserving. We are dealing with ECAs to illustrate the easiness of use of the decomposition theorem.

Let us note that if $d=1$ and $Q=\{0,1\}$, then conditions (S1) and (S2) of Definition \ref{split} collapse to
\[ 
h(1,0,0) + h(0,1,0) + h(0,0,1) = 1
 \quad \text{and} \quad   h(1,0,0),\, h(0,1,0),\, h(0,0,1) \in \{ 0,1\}.
\]
Thus, as mentioned in Example 1, there are only three  split functions: $h_1$, $h_2$, $h_3$ (see Table \ref{tab:binary1}), which all are local rules (\emph{i.e.}, their values belong to $\{0,1\}$).

\begin{table}[h]
	\centering
		\begin{tabular}{|c|c|c|c|c|c|c|c|c|l|}
		\hline
		 \rule{0pt}{1.2em} & 111 & 110 & 101 & 100 & 011 & 010 & 001 & 000 &  \\
		\hline
	     \rule{0pt}{1.2em}$h_1$ &	$1$ & $1$ & $1$ & $1$ & $0$ & $0$ & $0$ & $0$ & the shift-right rule (ECA 240)\\
		\hline
		 \rule{0pt}{1.2em}$h_2$ &	$1$ & $1$ & $0$ & $0$ & $1$ & $1$ & $0$ & $0$ & the identity rule (ECA 204)\\
		\hline
		 \rule{0pt}{1.2em}$h_3$ &	$1$ & $0$ & $1$ & $0$ & $1$ & $0$ & $1$ & $0$ & the shift-left rule (ECA 170)		\\
		\hline
		 \rule{0pt}{1.2em}$g_1$ &	$0$ & $-1$ & $0$ & $0$ & $1$ & $0$ & $0$ & $0$ & the basic perturbation\\
		 \hline
	     \rule{0pt}{1.2em}$h_1+ag_1$ &	$1$ & $1-a$ & $1$ & $1$ & $a$ & $0$ & $0$ & $0$ &\\
		\hline
		 \rule{0pt}{1.2em}$h_2+ag_1$ &	$1$ & $1-a$ & $0$ & $0$ & $1+a$ & $1$ & $0$ & $0$ &\\
		\hline
		 \rule{0pt}{1.2em}$h_3+ag_1$ &	$1$ & $-a$ & $1$ & $0$ & $1+a$ & $0$ & $1$ & $0$ &		\\
		\hline
	\end{tabular}
	\caption{The LUTs of all split functions $h_1$, $h_2$, $h_3$, the basic perturbation $g_1$ and local functions $h_i+ag_1$, for ECAs.} \label{tab:binary1}
\end{table}

On the other hand, the set $\CP$ contains only one pair $\lbrace \vv{0}{},\vv{v}{1}\rbrace$ (Example \ref{example:q-cd}), so, the perturbation space has dimension $1$, and as a basic perturbation we may take the local function $g_1$ presented in Table \ref{tab:binary1}. Thus, according to Theorem \ref{main-dec}, every number-conserving local rule can be written as $h_i + ag_1$, where $i\in\{1,2,3\}$ and $a\in\R$.

If the state set $Q$ is finite, each local function $f$ can be defined in tabular form by listing all elements of $\NN$ and the corresponding values of $f$. This kind of presentation will be referred to as the lookup table (LUT) of the local function $f$.

The LUTs of the local functions $h_i + ag_1$ are shown in Table \ref{tab:binary1}. In the case of $h_1$, only $a=0$ or $a=1$ give a local rule, otherwise $h_1$ yields a value that does not belong to $\{ 0,1\}$. In the case of $h_2$, only $a=0$ gives a local rule, while in the case of $h_3$, only $a=0$ and $a=-1$.

Finally, we get the five well-known number-conserving ECAs:
\begin{itemize}
\item $h_1$: the shift-right rule ECA 240,
\item $h_1 + g_1$: the traffic-right rule ECA 184,
\item $h_2$: the identity rule ECA 204,
\item $h_3$: the shift-left rule ECA 170,
\item $h_3 - g_1$: the traffic-left rule ECA 226.
\end{itemize}

\medskip

{\it Two-dimensional binary CAs} \\
First, let us recall that in the case $d=2$, the set $V$ has $5$ elements, namely $V=\{\vv{0}{},\vv{v}{1},\mbox{-}\vv{v}{1},\vv{v}{2},\mbox{-}\vv{v}{2}\}$ and that as a set $\CP$ we selected $\bigg\lbrace  
\lbrace \vv{0}{},\vv{v}{1}\rbrace,  \lbrace\vv{0}{},\vv{v}{2}\rbrace ,  \lbrace \vv{v}{1},\vv{v}{2}\rbrace, 
\lbrace\vv{v}{1},\mbox{-}\vv{v}{2}\rbrace \bigg\rbrace$. According to conditions (S1) and (S2) of Definition \ref{split},  we have to solve the following equation to find all split functions:
\begin{equation}\label{w2}
\ff{1}{0}{0}{0}{0}+\ff{0}{1}{0}{0}{0}+\ff{0}{0}{1}{0}{0}+\ff{0}{0}{0}{1}{0}+\ff{0}{0}{0}{0}{1} = 1,
\end{equation}
where $\ff{1}{0}{0}{0}{0},\;\ff{0}{1}{0}{0}{0},\;\ff{0}{0}{1}{0}{0},\;\ff{0}{0}{0}{1}{0},\;\ff{0}{0}{0}{0}{1}$ belong to $\{ 0,1\}$.
So we see that there are only $5$ split functions: $h_{\vv{0}{}}$, $h_{\vv{v}{1}}$, $h_{\mbox{-}\vv{v}{1}}$, $h_{\vv{v}{2}}$ and $h_{\mbox{-}\vv{v}{2}}$ (Table \ref{tab:binary-d2}), and, as in the one-dimensional case, all of them are local rules.

\begin{table}[h]
	\centering
		\begin{tabular}{|l|c|c|c|c|c|c|l|}
		\hline
		 \rule{0pt}{1.2em} & \setlength{\arraycolsep}{1pt}
\renewcommand{\arraystretch}{0.8}
$\begin{array}{rcl}
 & 1 &  \\
0 & 0 & 0 \\
 & 0 & 
\end{array}$ 
& 
$\setlength{\arraycolsep}{1pt}
\renewcommand{\arraystretch}{0.8}
\begin{array}{rcl}
 & 0 &  \\
1 & 0 & 0 \\
 & 0 & 
\end{array}$ 
& 
$\setlength{\arraycolsep}{1pt}
\renewcommand{\arraystretch}{0.8}
\begin{array}{rcl}
 & 0 &  \\
0 & 1 & 0 \\
 & 0 & 
\end{array}$ 
& 
$\setlength{\arraycolsep}{1pt}
\renewcommand{\arraystretch}{0.8}
\begin{array}{rcl}
 & 0 &  \\
0 & 0 & 1 \\
 & 0 & 
\end{array}$ 
& 
$\setlength{\arraycolsep}{1pt}
\renewcommand{\arraystretch}{0.8}
\begin{array}{rcl}
 & 0 &  \\
0 & 0 & 0 \\
 & 1 & 
\end{array}$ 
& 
$\setlength{\arraycolsep}{1pt}
\renewcommand{\arraystretch}{0.8}
\begin{array}{rcl}
 & q_1 &  \\
q_2 & q_3 & q_4 \\
 & q_5 & 
\end{array}$ &
 \\
\hline
 \rule{0pt}{1.2em}$h_{\vv{0}{}}$ &	0 & 0 & 1 & 0 & 0 & $q_3$ & the identity rule\\
\hline
 \rule{0pt}{1.2em}$h_{\vv{v}{1}}$ &	0 & 0 & 0 & 1 & 0 & $q_4$  & the shift-left rule\\
\hline
 \rule{0pt}{1.2em}$h_{\mbox{-}\vv{v}{1}}$ &	0 & 1 & 0 & 0 & 0 & $q_2$ & the shift-right rule 		\\
\hline
\rule{0pt}{1.2em}$h_{\vv{v}{2}}$ &	1 & 0 & 0 & 0 & 0 & $q_1$ & the shift-down rule \\
\hline
\rule{0pt}{1.2em}$h_{\mbox{-}\vv{v}{2}}$ &	0 & 0 & 0 & 0 & 1 & $q_5$ & the shift-up rule 		\\
\hline
	\end{tabular}
	\caption{All the split functions for two-dimensional binary CAs.}\label{tab:binary-d2}
\end{table}

Recalling Example~\ref{example:q-cd}, we know that the perturbation space has dimension $4$, and as basic perturbations we may take the local functions $g_1$, $g_2$, $g_3$, $g_4$ presented in Table \ref{tab:binary-d2per}. Thus, according to Theorem \ref{main-dec}, every number-conserving local rule can be represented as $h_{\vv{v}{}} + ag_1+bg_2+cg_3+dg_4$, where $\vv{v}{}\in V$ and $a,b,c,d\in\R$. 
As $h_{\mbox{-}\vv{v}{1}}$, $h_{\vv{v}{2}}$, $h_{\mbox{-}\vv{v}{2}}$ are equivalent with
$h_{\vv{v}{1}}$ by rotation, we restrict our discussion to $h_{\vv{0}{}}$  and $h_{\vv{v}{1}}$.
The LUTs of the local functions $h_{\vv{0}{}} + ag_1+bg_2+cg_3+dg_4$ and $h_{\vv{v}{1}} + ag_1+bg_2+cg_3+dg_4$  are shown in Table \ref{tab:binary-d2per}.

\begin{table}[p]
	\centering
	\begin{tabular}{||c||c||c|c||c|c|c|c||l|l|}
		\hline
		 & $l_{i}$ & $h_{\vv{0}{}}$ & $h_{\vv{v}{1}}$  & $g_1$ & $g_2$ & $g_3$ & $g_4$ & $f_0$ & $f_1$ \\
		 \hline
		 \hline
		\setlength{\arraycolsep}{1pt}
\renewcommand{\arraystretch}{0.8}
\scriptsize{$\begin{array}{rcl}
 & 0 &  \\
0 & 0 & 0 \\
 & 0 & 
\end{array}$} 
& $l_0$  & 0 & 0 & & & & & &
 \\
	\hline
	\setlength{\arraycolsep}{1pt}
\renewcommand{\arraystretch}{0.8}
\scriptsize{$\begin{array}{rcl}
 & 0 &  \\
0 & 0 & 0 \\
 & 1 & 
\end{array}$} 
& $l_{1}$ & 0 & 0 &  & & & & &
 \\
	\hline
	\setlength{\arraycolsep}{1pt}
\renewcommand{\arraystretch}{0.8}
\scriptsize{$\begin{array}{rcl}
 & 0 &  \\
0 & 0 & 1 \\
 & 0 & 
\end{array}$} 
& $l_{2}$ & 0 & 1  & & & & & & 1
 \\
	\hline
	\setlength{\arraycolsep}{1pt}
\renewcommand{\arraystretch}{0.8}
\scriptsize{$\begin{array}{rcl}
 & 0 &  \\
0 & 0 & 1 \\
 & 1 & 
\end{array}$} 
& $l_{3}$ & 0 & 1 &  & & & 1 & $d$ & 1+$d$
 \\
	\hline
	\setlength{\arraycolsep}{1pt}
\renewcommand{\arraystretch}{0.8}
\scriptsize{$\begin{array}{rcl}
 & 0 &  \\
0 & 1 & 0 \\
 & 0 & 
\end{array}$} 
& $l_{4}$ & 1 & 0  & & & & & 1 & 
 \\
	\hline
	\setlength{\arraycolsep}{1pt}
\renewcommand{\arraystretch}{0.8}
\scriptsize{$\begin{array}{rcl}
 & 0 &  \\
0 & 1 & 0 \\
 & 1 & 
\end{array}$} 
& $l_{5}$ & 1 & 0 &  & $-1$ & & & $1-b$ & $-b$
 \\
	\hline
	\setlength{\arraycolsep}{1pt}
\renewcommand{\arraystretch}{0.8}
\scriptsize{$\begin{array}{rcl}
 & 0 &  \\
0 & 1 & 1 \\
 & 0 & 
\end{array}$} 
& $l_{6}$ & 1 & 1 &  $1$ & & & & 1+ $a$ & 1+ $a$
 \\
	\hline
	\setlength{\arraycolsep}{1pt}
\renewcommand{\arraystretch}{0.8}
\scriptsize{$\begin{array}{rcl}
 & 0 &  \\
0 & 1 & 1 \\
 & 1 & 
\end{array}$} 
& $l_{7}$ & 1 & 1 &  $1$ & $-1$ & & $1$ & 1+$a-b+d$ &  1+$a-b+d$
 \\
	\hline
	\setlength{\arraycolsep}{1pt}
\renewcommand{\arraystretch}{0.8}
\scriptsize{$\begin{array}{rcl}
 & 0 &  \\
1 & 0 & 0 \\
 & 0 & 
\end{array}$} 
& $l_{8}$ & 0 & 0 &  & & & & &
 \\
	\hline
	\setlength{\arraycolsep}{1pt}
\renewcommand{\arraystretch}{0.8}
\scriptsize{$\begin{array}{rcl}
 & 0 &  \\
1 & 0 & 0 \\
 & 1 & 
\end{array}$} 
& $l_{9}$ & 0 & 0 &  & & $-1$ & & $-c$ & $-c$
 \\
	\hline
	\setlength{\arraycolsep}{1pt}
\renewcommand{\arraystretch}{0.8}
\scriptsize{$\begin{array}{rcl}
 & 0 &  \\
1 & 0 & 1 \\
 & 0 & 
\end{array}$} 
& $l_{10}$ & 0 & 1 &  & & & & & 1
 \\
	\hline
	\setlength{\arraycolsep}{1pt}
\renewcommand{\arraystretch}{0.8}
\scriptsize{$\begin{array}{rcl}
 & 0 &  \\
1 & 0 & 1 \\
 & 1 & 
\end{array}$} 
& $l_{11}$ & 0 & 1 &  & & $-1$ & $1$ & $-c+d$ & $1-c+d$
 \\
	\hline
	\setlength{\arraycolsep}{1pt}
\renewcommand{\arraystretch}{0.8}
\scriptsize{$\begin{array}{rcl}
 & 0 &  \\
1 & 1 & 0 \\
 & 0 & 
\end{array}$} 
& $l_{12}$ & 1 & 0 &  $-1$ & & & & $1-a$ & $-a$
 \\
	\hline
	\setlength{\arraycolsep}{1pt}
\renewcommand{\arraystretch}{0.8}
\scriptsize{$\begin{array}{rcl}
 & 0 &  \\
1 & 1 & 0 \\
 & 1 & 
\end{array}$} 
& $l_{13}$ & 1 & 0 &  $-1$ & $-1$ & $-1$ & & $1-a-b-c$ & $-a-b-c$
 \\
	\hline
	\setlength{\arraycolsep}{1pt}
\renewcommand{\arraystretch}{0.8}
\scriptsize{$\begin{array}{rcl}
 & 0 &  \\
1 & 1 & 1 \\
 & 0 & 
\end{array}$} 
& $l_{14}$ & 1 & 1 &  & & & & 1 & 1
 \\
	\hline
	\setlength{\arraycolsep}{1pt}
\renewcommand{\arraystretch}{0.8}
\scriptsize{$\begin{array}{rcl}
 & 0 &  \\
1 & 1 & 1 \\
 & 1 & 
\end{array}$} 
& $l_{15}$ & 1 & 1 &  & $-1$ & $-1$ & $1$ & $1-b-c+d$ & $1-b-c+d$ 
 \\
	\hline
	\setlength{\arraycolsep}{1pt}
\renewcommand{\arraystretch}{0.8}
\scriptsize{$\begin{array}{rcl}
 & 1 &  \\
0 & 0 & 0 \\
 & 0 & 
\end{array}$} 
& $l_{16}$ & 0 & 0 &  & & & & &
 \\
	\hline
	\setlength{\arraycolsep}{1pt}
\renewcommand{\arraystretch}{0.8}
\scriptsize{$\begin{array}{rcl}
 & 1 &  \\
0 & 0 & 0 \\
 & 1 & 
\end{array}$} 
& $l_{17}$ & 0 & 0 &  & & & & &
 \\
	\hline
	\setlength{\arraycolsep}{1pt}
\renewcommand{\arraystretch}{0.8}
\scriptsize{$\begin{array}{rcl}
 & 1 &  \\
0 & 0 & 1 \\
 & 0 & 
\end{array}$} 
& $l_{18}$ & 0 & 1 &  & & $1$ & & $c$ & $1+c$
 \\
	\hline
	\setlength{\arraycolsep}{1pt}
\renewcommand{\arraystretch}{0.8}
\scriptsize{$\begin{array}{rcl}
 & 1 &  \\
0 & 0 & 1 \\
 & 1 & 
\end{array}$} 
& $l_{19}$ & 0 & 1 &  & & $1$ & $1$ & $c+d$ & $1+c+d$
 \\
	\hline
	\setlength{\arraycolsep}{1pt}
\renewcommand{\arraystretch}{0.8}
\scriptsize{$\begin{array}{rcl}
 & 1 &  \\
0 & 1 & 0 \\
 & 0 & 
\end{array}$} 
& $l_{20}$ & 1 & 0 &  & $1$ & & & $1+b$ & $b$
 \\
	\hline
	\setlength{\arraycolsep}{1pt}
\renewcommand{\arraystretch}{0.8}
\scriptsize{$\begin{array}{rcl}
 & 1 &  \\
0 & 1 & 0 \\
 & 1 & 
\end{array}$} 
& $l_{21}$ & 1 & 0 &  & & & & 1 &
 \\
	\hline
	\setlength{\arraycolsep}{1pt}
\renewcommand{\arraystretch}{0.8}
\scriptsize{$\begin{array}{rcl}
 & 1 &  \\
0 & 1 & 1 \\
 & 0 & 
\end{array}$} 
& $l_{22}$ & 1 & 1 &  $1$ & $1$ & $1$ & & $1+a+b+c$ & $1+a+b+c$
 \\
	\hline
	\setlength{\arraycolsep}{1pt}
\renewcommand{\arraystretch}{0.8}
\scriptsize{$\begin{array}{rcl}
 & 1 &  \\
0 & 1 & 1 \\
 & 1 & 
\end{array}$} 
& $l_{23}$ & 1 & 1 &  $1$ & & $1$ & $1$ & $1+a+c+d$ & $1+a+c+d$
 \\
	\hline
	\setlength{\arraycolsep}{1pt}
\renewcommand{\arraystretch}{0.8}
\scriptsize{$\begin{array}{rcl}
 & 1 &  \\
1 & 0 & 0 \\
 & 0 & 
\end{array}$} 
& $l_{24}$  & 0 & 0 &  & & & $-1$ & $-d$ & $-d$
 \\
	\hline
	\setlength{\arraycolsep}{1pt}
\renewcommand{\arraystretch}{0.8}
\scriptsize{$\begin{array}{rcl}
 & 1 &  \\
1 & 0 & 0 \\
 & 1 & 
\end{array}$} 
& $l_{25}$ & 0 & 0 &  & & $-1$ & $-1$ & $-c-d$ & $-c-d$
 \\
	\hline
	\setlength{\arraycolsep}{1pt}
\renewcommand{\arraystretch}{0.8}
\scriptsize{$\begin{array}{rcl}
 & 1 &  \\
1 & 0 & 1 \\
 & 0 & 
\end{array}$} 
& $l_{26}$ & 0 & 1 &  & & $1$ & $-1$ & $c-d$ & $1+c-d$
 \\
	\hline
	\setlength{\arraycolsep}{1pt}
\renewcommand{\arraystretch}{0.8}
\scriptsize{$\begin{array}{rcl}
 & 1 &  \\
1 & 0 & 1 \\
 & 1 & 
\end{array}$} 
& $l_{27}$ & 0 & 1 &  & & & & & 1
 \\
	\hline
	\setlength{\arraycolsep}{1pt}
\renewcommand{\arraystretch}{0.8}
\scriptsize{$\begin{array}{rcl}
 & 1 &  \\
1 & 1 & 0 \\
 & 0 & 
\end{array}$} 
& $l_{28}$ & 1 & 0 &  $-1$ & $1$ & & $-1$ & $1-a+b-d$ & $-a+b-d$
 \\
	\hline
	\setlength{\arraycolsep}{1pt}
\renewcommand{\arraystretch}{0.8}
\scriptsize{$\begin{array}{rcl}
 & 1 &  \\
1 & 1 & 0 \\
 & 1 & 
\end{array}$} 
& $l_{29}$ & 1 & 0 &  $-1$ & & $-1$ & $-1$ & $1-a-c-d$ & $-a-c-d$
 \\
	\hline
	\setlength{\arraycolsep}{1pt}
\renewcommand{\arraystretch}{0.8}
\scriptsize{$\begin{array}{rcl}
 & 1 &  \\
1 & 1 & 1 \\
 & 0 & 
\end{array}$} 
& $l_{30}$ & 1 & 1 &  & $1$ & $1$ & $-1$ & $1+b+c-d$ & $1+b+c-d$
 \\
	\hline
	\setlength{\arraycolsep}{1pt}
\renewcommand{\arraystretch}{0.8}
\scriptsize{$\begin{array}{rcl}
 & 1 &  \\
1 & 1 & 1 \\
 & 1 & 
\end{array}$} 
& $l_{31}$ & 1 & 1 &  & & &  & 1 & 1
 \\
	\hline
	\end{tabular}
    \caption{The LUTs of the split functions $h_{\vv{0}{}}$, $h_{\vv{v}{1}}$, basic perturbations and local functions $f_0 = h_{\vv{0}{}}+ag_1 + bg_2 + cg_3 + dg_4$ and $f_1 = h_{\vv{v}{1}}+ag_1 + bg_2 + cg_3 + dg_4$, for two-dimensional binary CAs.}\label{tab:binary-d2per}
\end{table}

First, let us consider the local function $f_0=h_{\vv{0}{}} + ag_1+bg_2+cg_3+dg_4$. Comparing $l_6 = 1+a$ and $l_{12} = 1-a$, we deduce that if $f_0$ would be a local rule, then it should hold that $a=0$. Analogously, we get $b=0$ (comparing $l_5 = 1-b$ and $l_{20}= 1+b$), $c=0$ (comparing $l_9= -c$ and $l_{18}= c$) and $d=0$ (comparing $l_3= d$ and $l_{24}= -d$). 
So, finally, $f_0$ is a local rule if and only if $a=b=c=d=0$.

Now, let us consider the local function $f_1=h_{\vv{v}{1}} + ag_1+bg_2+cg_3+dg_4$. Comparing $l_5 = -b$ and $l_{20} = b$, we find that $b=0$. As $l_{11}=1-c-d$ and $l_{26}=1+c-d$, it must hold that $c=d$. On the other hand, since $l_3=1+d$ and $l_{24}=-d$, it must hold that $d=-1$ or $d=0$. Analogously, from $l_9=-c$ and $l_{18}=1+c$, we deduce that $c=-1$ or $c=0$. 
Thus, we have $c=d=-1$ or $c=d=0$. But as $l_{25}=-c-d$, the former is impossible. Now, putting $b=c=d=0$ into the LUT of $f_1$, we see that the conditions on $a$ are: $1+a,-a\in\{0,1\}$, so $a=0$ or $a=-1$. 
Summarizing, there are two local rules of the form $h_{\vv{v}{1}} + ag_1+bg_2+cg_3+dg_4$:
\begin{itemize}
\item $h_{\vv{v}{1}}$: the shift-left rule,
\item $h_{\vv{v}{1}} - g_1$: the traffic-left rule.
\end{itemize}

Consequently, we get that there are only 9 number-conserving local rules in the case of two-dimensional binary CAs: the identity rule, and the shift and traffic rules in each of the four directions (right, left, up and down).

\medskip

{\it Three-dimensional binary CAs} \\
Reasoning in a similar way as in the case of one- or two-dimensional number-conserving binary $CAs$, we obtain $7$ split functions and 13 number-conserving local rules: the identity rule, and the shift and traffic rules in each of the six directions.

\medskip

{\it The case $d>3$} \\
We conjecture that for any $d$ there are no non-trivial number-conserving binary CAs.
\begin{con}\label{binary}
Let the dimension $d\geq 1$ be given. There are exactly $4d+1$ number-conserving binary CAs with the von Neumann neighborhood: the identity rule, and the shift and traffic rules in each of the $2d$ possible directions.
\end{con}

\subsection{Number-conserving three-state CAs}

{\it One-dimensional three-state CAs} \\
Here, we consider the state set $Q=\{0,1,2\}$ in the case of one-dimensional CAs, for which Boccara and Fukś~\cite{BoccaraF02} found all number-conserving CAs. Yet, our approach allows us to do it without the use of a computer.

Conditions (S1) and (S2) of Definition \ref{split} now simplify to
\[ 
h(1,0,0) + h(0,1,0) + h(0,0,1) = 1
\]
and
\[ 
h(2,0,0) + h(0,2,0) + h(0,0,2) = 2,
\] 
with $h(1,0,0),\, h(0,1,0),\, h(0,0,1),\, h(2,0,0),\, h(0,2,0),\, h(0,0,2) \in \{ 0,1,2\}$.
Thus, as mentioned in Example \ref{example:q}, there are exactly $\binom{3}{1} \binom{4}{2}=18$ split functions, ten of which, namely $h_0$,  $h_1$, $\ldots$, $h_9$ are presented in Table \ref{tab:three-d1per}. The remaining eight $h_{10}$, $h_{11}$, $\ldots$, $h_{17}$ are reflections of $h_1$, $h_2$,  $\ldots$, $h_8$. Note that not all of them are local rules.

Since $d=1$,  the set $\CP$ contains only one pair: $\CP=\big\lbrace  \lbrace \vv{0}{},\vv{v}{1}\rbrace \big\rbrace$. 
Moreover, from Example \ref{example:q-cd}, we have that the perturbation space has dimension $4$, since to define a perturbation, we have to set the values it maps to for dimers from $\D = \left\{ \dimer{\vv{0}{}:1}{\vv{v}{1}:1},\dimer{\vv{0}{}:1}{\vv{v}{1}:2}, \dimer{\vv{0}{}:2}{\vv{v}{1}:1}, \dimer{\vv{0}{}:2}{\vv{v}{1}:2} \right\} = \left\{ 011, 012,021,022 \right\}$.
Thus, as basic perturbations we may take the local functions $g_1$, $g_2$, $g_3$, $g_4$, where $g_k$ maps to $1$ on the $k$-th dimer listed in $\D$ and to $0$ on the remaining dimers from $\D$. According to Theorem \ref{main-dec}, every number-conserving local rule can be written as $h_i + ag_1+bg_2+cg_3+dg_4$, where $i\in\{0,1,\ldots,17\}$ and $a,b,c,d\in\R$.

In Table \ref{tab:three-d1per}, we show the LUTs of all basic perturbations and the form of the LUT of a general one. 
Now, all we have to check is whether $h_i + ag_1+bg_2+cg_3+dg_4$ is a local rule, \emph{i.e.}, it yields values in $Q=\{0,1,2\}$.

\begin{table}[t]
	\centering
	\begin{tabular}{||c||c||c|c|c|c|c|c|c|c|c|c||c|c|c|c||r||}		
    \hline
		 & $l_{i}$ & $h_{0}$ & $h_{1}$ & $h_{2}$ & $h_{3}$ & $h_{4}$ & $h_{5}$ & $h_{6}$ & $h_{7}$ & $h_{8}$ & $h_{9}$  & $g_1$ & $g_2$ & $g_3$ & $g_4$ & $g$  \\
		 \hline
		 \hline
	000
& $l_{0}$  & 0 & 0 & 0 & 0 & 0 & 0 & 0 & 0 & 0 & 0 &  &  &  &  & 
 \\
	\hline
	001
& $l_{1}$  & 0 & 0 & 0 & 1 & 1 & 1 & 1 & 1 & 1 & 0 &  &  &  &  & 
 \\
	\hline
	002
& $l_{2}$  & 0 & 2 & 1 & 2 & 0 & 0 & 1 & 1 & 0 & 1 &  &  &  &  & 
 \\
	\hline
	010
& $l_{3}$  & 1 & 1 & 1 & 0 & 0 & 0 & 0 & 0 & 0 & 1 &  &  &  &  & 
 \\
	\hline
	011
& $l_{4}$  & 1 & 1 & 1 & 1 & 1 & 1 & 1 & 1 & 1 & 1 & 1 &  &  &  & $a$ 
 \\
	\hline
	012
& $l_{5}$  & 1 & 3 & 2 & 2 & 0 & 0 & 1 & 1 & 0 & 2 &  & 1 &  &  & $b$
 \\
	\hline
	020
& $l_{6}$  & 2 & 0 & 1 & 0 & 2 & 0 & 1 & 0 & 1 & 0 &  &  &  &  & 
 \\
	\hline
	021
& $l_{7}$  & 2 & 0 & 1 & 1 & 3 & 1 & 2 & 1 & 2 & 0 &  &  & 1 &  & $c$
 \\
	\hline
	022
& $l_{8}$  & 2 & 2 & 2 & 2 & 2 & 0 & 2 & 1 & 1 & 1 &  &  &  & 1 & $d$
 \\
	\hline
	100
& $l_{9}$  & 0 & 0 & 0 & 0 & 0 & 0 & 0 & 0 & 0 & 0 &  &  &  &  & 
 \\
	\hline
	101
& $l_{10}$  & 0 & 0 & 0 & 1 & 1 & 1 & 1 & 1 & 1 & 0 &  &  &  &  & 
 \\
	\hline
	102
& $l_{11}$  & 0 & 2 & 1 & 2 & 0 & 0 & 1 & 1 & 0 & 1 &  &  &  &  & 
 \\
	\hline
	110
& $l_{12}$  & 1 & 1 & 1 & 0 & 0 & 0 & 0 & 0 & 0 & 1 & -1 &  &  &  & $-a$
 \\
	\hline
	111
& $l_{13}$  & 1 & 1 & 1 & 1 & 1 & 1 & 1 & 1 & 1 & 1 &  &  &  &  & 
 \\
	\hline
	112
& $l_{14}$  & 1 & 3 & 2 & 2 & 0 & 0 & 1 & 1 & 0 & 2 & -1 & 1 &  &  & $-a+b$
 \\
	\hline
	120
& $l_{15}$  & 2 & 0 & 1 & 0 & 2 & 0 & 1 & 0 & 1 & 0 &  & -1 &  &  & $-b$
 \\
	\hline
	121
& $l_{16}$  & 2 & 0 & 1 & 1 & 3 & 1 & 2 & 1 & 2 & 0 &  & -1 & 1 &  & $-b+c$
 \\
	\hline
	122
& $l_{17}$  & 2 & 2 & 2 & 2 & 2 & 0 & 2 & 1 & 1 & 1 &  & -1 &  & 1 & $-b+d$
 \\
	\hline
	200
& $l_{18}$  & 0 & 0 & 0 & 0 & 0 & 2 & 0 & 1 & 1 & 1 &  &  &  &  & 
 \\
	\hline
	201
& $l_{19}$  & 0 & 0 & 0 & 1 & 1 & 3 & 1 & 2 & 2 & 1 &  &  &  &  & 
 \\
	\hline
	202
& $l_{20}$  & 0 & 2 & 1 & 2 & 0 & 2 & 1 & 2 & 1 & 2 &  &  &  &  & 
 \\
	\hline
	210
& $l_{21}$  & 1 & 1 & 1 & 0 & 0 & 2 & 0 & 1 & 1 & 2 &  &  & -1 &  & $-c$
 \\
	\hline
	211
& $l_{22}$  & 1 & 1 & 1 & 1 & 1 & 3 & 1 & 2 & 2 & 2 & 1 &  & -1 &  & $a-c$
 \\
	\hline
	212
& $l_{23}$  & 1 & 3 & 2 & 2 & 0 & 2 & 1 & 2 & 1 & 3 &  & 1 & -1 &  & $b-c$
 \\
	\hline
	220
& $l_{24}$  & 2 & 0 & 1 & 0 & 2 & 2 & 1 & 1 & 2 & 1 &  &  &  & -1 & $-d$
 \\
	\hline
	221
& $l_{25}$  & 2 & 0 & 1 & 1 & 3 & 3 & 2 & 2 & 3 & 1 &  &  & 1 & -1 & $c-d$
 \\
	\hline
	222
& $l_{26}$  & 2 & 2 & 2 & 2 & 2 & 2 & 2 & 2 & 2 & 2 &  &  &  &  & 
 \\
	\hline
	\end{tabular}
		\caption{The LUTs of the split functions $h_0$,  $h_1$, $\ldots$, $h_9$, the basic perturbations and a general perturbation $g=ag_1 + bg_2 + cg_3 + dg_4$ in the case of one-dimensional three-state CAs.}\label{tab:three-d1per}
\end{table}

For that propose, we have to consider every $h_i$, $i=0,1,\ldots,9$ separately.
For example, if $i=0$, the problem boils down to solving the following system of inequalities in $Q=\{0,1,2\}$:
\[
\left\{
\begin{array}{lclcl}
		0\leq 1+a\leq 2		&\; & 0\leq 1-a+b\leq 2 &\; & 0\leq 1+a-c\leq 2\\
		0\leq 1+b\leq 2		&\; & 0\leq 2-b\leq 2 &\; & 0\leq 1+b-c\leq 2\\
		0\leq 2+c\leq 2		&\; & 0\leq 2-b+c\leq 2 &\; & 0\leq 2-d\leq 2\\
		0\leq 2+d\leq 2		&\; & 0\leq 2-b+d\leq 2 &\; & 0\leq 2+c-d\leq 2\\
		0\leq 1-a\leq 2		&\; & 0\leq 1-c\leq 2 &\; & 
\end{array}\;,
\right.
\]
which can easily be reduced to $a\in\{-1,0,1\}$, $b\in\{0,1\}$, $c\in\{-1,0\}$, $d=0$, and
\[
\left\{
\begin{array}{l}
		-1\leq a-b\leq 1	 \\
		-1\leq c-a\leq 1	 \\
		0\leq b-c\leq 1	 
\end{array}\;.
\right.
\]
So, for $h_0$ (the identity rule), we have seven solutions for $(a,b,c,d)$: $(-1,0,0,0)$, $(-1,0,-1,0)$, $(0,0,-1,0)$, $(0,0,0,0)$, $(0,1,0,0)$, $(1,0,0,0)$ and $(1,1,0,0)$.

For $h_3$ (the shift-left rule), we get the following system of inequalities:
\[
\left\{
\begin{array}{lclcl}
		0\leq 1+a\leq 2		&\; & 0\leq 2-a+b\leq 2 &\; & 0\leq 1+a-c\leq 2\\
		0\leq 2+b\leq 2		&\; & 0\leq -b\leq 2 &\; & 0\leq 2+b-c\leq 2\\
		0\leq 1+c\leq 2		&\; & 0\leq 1-b+c\leq 2 &\; & 0\leq -d\leq 2\\
		0\leq 2+d\leq 2		&\; & 0\leq 2-b+d\leq 2 &\; & 0\leq 1+c-d\leq 2\\
		0\leq -a\leq 2		&\; & 0\leq -c\leq 2 &\; & 
\end{array}\;,
\right.
\]
which reduces to $a\in\{-1,0\}$, $b\in\{-2,-1,0\}$, $c\in\{-1,0\}$, $d\in\{-2,-1,0\}$ and
\[
\left\{
\begin{array}{l}
		0\leq a-b\leq 2	 \\
		-1\leq c-a\leq 1	 \\
		-1\leq c-d\leq 1	 \\
			0\leq b-d\leq 2	 \\
		0\leq c-b\leq 1	 
\end{array}\;.
\right.
\]
Thus, now we have 10 solutions: $(0,-2,-1,-2)$, $(0,-1,-1,-2)$, $(0,-1,-1,-1)$, $(0,-1,0,-1)$,  $(-1,-2,-1,-2)$,  $(-1,-1,-1,-2)$, $(-1,-1,0,-1)$, $(0,0,0,-1)$, $(-1,-1,-1,-1)$, $(0,0,0,0)$. 

For $h_{5}$, we see at once that there is no solution at all, as for any perturbation we have $l_{19}=3\not\in\{0,1,2\}$.

One can proceed similarly in the seven other cases. In this way one arrives at $7$, $9$, $11$, $10$, $5$, $0$, $10$, $10$, $6$, $15$ solutions for $h_0$, $h_1$, $h_2$, $h_3$, $h_4$, $h_5$, $h_6$, $h_7$, $h_8$, $h_9$, respectively.
The solutions for $h_{11}$,..., $h_{17}$ can be obtained from the solutions of their reflection $h_1$,..., $h_8$. 

So, we have $7 + 2(9+11+10+5+0+10+10+6) + 15 = 144$ one-dimensional three-state number-conserving local rules, which agrees with the findings of~\cite{BoccaraF02}.

\medskip

{\it Two-dimensional three-state CAs
} \\
Now, conditions (S1) and (S2) of Definition \ref{split}, as mentioned in Example \ref{example:q}, give exactly $\binom{5}{1}\binom{6}{2}=75$ split functions. Thus it is still possible to consider every split function separately and calculate all number-conserving local rules connected with it without the use of a computer. Moreover, we can exploit eight symmetries of the von Neumann neighborhood in the two-dimensional space, so, there are only 16 significantly different split functions. As a result, we get that there are 1327 number-conserving local rules, which agrees with the findings of~\cite{enum}.

\medskip

{\it Three-dimensional three-state CAs
} \\
So far, there were no results about three-dimensional three-state number-conserving CAs with the von Neumann neighborhood. This is due to the fact that we are looking in a huge set of all local rules of cardinality $3^{3^7}\approx 2.9\cdot 10^{1043}$. Now, using the decomposition theorem we can find all three-dimensional three-state number-conserving CAs with the von Neumann neighborhood.

More precisely, conditions (S1) and (S2) of Definition \ref{split} simplify to
\begin{equation}
\label{split1}
\fff{1}{0}{0}{0}{0}{0}{0} +\fff{0}{0}{0}{0}{0}{0}{1} +\fff{0}{1}{0}{0}{0}{0}{0} +\fff{0}{0}{1}{0}{0}{0}{0} +\fff{0}{0}{0}{1}{0}{0}{0} +\fff{0}{0}{0}{0}{0}{1}{0} +\fff{0}{0}{0}{0}{1}{0}{0}   = 1
\end{equation}
and
\begin{equation}
\label{split2}
\fff{2}{0}{0}{0}{0}{0}{0} +\fff{0}{0}{0}{0}{0}{0}{2} +\fff{0}{2}{0}{0}{0}{0}{0} +\fff{0}{0}{2}{0}{0}{0}{0} +\fff{0}{0}{0}{2}{0}{0}{0} +\fff{0}{0}{0}{0}{0}{2}{0} +\fff{0}{0}{0}{0}{2}{0}{0}   = 2.
\end{equation}
Thus, there are exactly $\binom{7}{1}\binom{8}{2}=196$ split functions. We number them by giving two strings of digits: $h^{(1)}$ and $h^{(2)}$ -- an ordered list of the terms of the sum in the left side of Eq.~(\ref{split1}) and Eq.~(\ref{split2}), respectively.

From Example \ref{example:q-cd}, we know that the perturbation space has dimension $36$ and to define a perturbation, we need to set the values it maps to for dimers from 
\[
\D = \left\{ \dimer{\vv{u}{}:p}{\vv{w}{}:q} \mid  \lbrace\vv{u}{},\vv{w}{}\rbrace\in\CP,\; p, q\in \{1,2\}\right\}\,,
\]
where we may choose, for example,
\[
\CP=\bigg\lbrace
\lbrace\vv{0}{},\vv{v}{1}\rbrace, 
\lbrace\vv{v}{1},\mbox{-}\vv{v}{2}\rbrace, \lbrace\vv{v}{1},\vv{v}{3}\rbrace,  \lbrace\vv{v}{1},\mbox{-}\vv{v}{3}\rbrace,    
  \lbrace\vv{0}{},\mbox{-}\vv{v}{2}\rbrace,  \lbrace\vv{0}{},\mbox{-}\vv{v}{3}\rbrace,  \lbrace\mbox{-}\vv{v}{1},\mbox{-}\vv{v}{2}\rbrace,  \lbrace\mbox{-}\vv{v}{2},\mbox{-}\vv{v}{3}\rbrace,  \lbrace\mbox{-}\vv{v}{2},\vv{v}{3}\rbrace
\bigg\rbrace\,.
\]Thus, as basic perturbations we may take the perturbations $g_{\vv{u}{}:p,\vv{w}{}:q}$, for $\lbrace\vv{u}{},\vv{w}{}\rbrace\in\CP,\; p, q\in \{1,2\}$ -- a given perturbation $g_{\vv{u}{}:p,\vv{w}{}:q}$ maps to $1$ on the dimer $\dimer{\vv{u}{}:p}{\vv{w}{}:q}$ and $0$ on the remaining dimers from~$\D$. Recall that it is zero on monomers too, and on other $N\in\NN$ is given by (\ref{eq:thm-pert}). 
Now, according to Theorem \ref{main-dec}, for every split function $h$ (given by setting $h^{(1)}$ and $h^{(2)}$), it is sufficient to check for which integers 
$a_{\vv{u}{}:p,\vv{w}{}:q}$ the local function
\begin{equation}\label{evaluation}
f= h + \sum_{\lbrace\vv{u}{},\vv{w}{}\rbrace\in\CP}\left( \sum_{p,q\in \{1,2\}} a_{\vv{u}{}:p,\vv{w}{}:q} g_{\vv{u}{}:p,\vv{w}{}:q}\right)
\end{equation}
is a local rule, \emph{i.e.}, its values do not go outside $Q$.

As for any dimer $\dimer{\vv{u}{0}:p_0}{\vv{w}{0}:q_0}$ from $\D$ it holds that
\[
f(\dimer{\vv{u}{0}:p_0}{\vv{w}{0}:q_0}) = h(\dimer{\vv{u}{0}:p_0}{\vv{w}{0}:q_0}) + \sum_{\lbrace\vv{u}{},\vv{w}{}\rbrace\in\CP}\left( \sum_{p,q\in \{1,2\}} a_{\vv{u}{}:p,\vv{w}{}:q} g_{\vv{u}{}:p,\vv{w}{}:q}(\dimer{\vv{u}{0}:p_0}{\vv{w}{0}:q_0})\right)
\]
\[
 = h(M_{\vv{u}{0}:p_0}) + h(M_{\vv{w}{0}:q_0}) +  a_{\vv{u}{0}:p_0,\vv{w}{0}:q_0} \in Q\; , 
\]
we obtain the following bounds for the integer $a_{\vv{u}{0}:p_0,\vv{w}{0}:q_0}$:
\begin{equation}\label{estimation1}
-\Big( h(M_{\vv{u}{0}:p_0}) + h(M_{\vv{w}{0}:q_0})\Big) \leq  a_{\vv{u}{0}:p_0,\vv{w}{0}:q_0}
\leq 2- \Big( h(M_{\vv{u}{0}:p_0}) + h(M_{\vv{w}{0}:q_0})\Big)\; .
\end{equation}
On the other hand, using the matching dimer $\dimer{\vv{\mbox{-}w}{0}:p_0}{\vv{\mbox{-}u}{0}:q_0}$, we get
\[
f(\dimer{\vv{\mbox{-}w}{0}:p_0}{\vv{\mbox{-}u}{0}:q_0}) = h(\dimer{\vv{\mbox{-}w}{0}:p_0}{\vv{\mbox{-}u}{0}:q_0}) + \sum_{\lbrace\vv{u}{},\vv{w}{}\rbrace\in\CP}\left( \sum_{p,q\in \{1,2\}} a_{\vv{u}{}:p,\vv{w}{}:q} g_{\vv{u}{}:p,\vv{w}{}:q}(\dimer{\vv{\mbox{-}w}{0}:p_0}{\vv{\mbox{-}u}{0}:q_0})\right)
\]
\[
 = h(M_{\vv{\mbox{-}w}{0}:p_0}) + h(M_{\vv{\mbox{-}u}{0}:q_0}) -  a_{\vv{u}{0}:p_0,\vv{w}{0}:q_0} \in Q\; , 
\]
thus 
\begin{equation}\label{estimation2}
 h(M_{\vv{\mbox{-}w}{0}:p_0}) + h(M_{\vv{\mbox{-}u}{0}:q_0}) -2 \leq  a_{\vv{u}{0}:p_0,\vv{w}{0}:q_0}
\leq  h(M_{\vv{\mbox{-}w}{0}:p_0}) + h(M_{\vv{\mbox{-}u}{0}:q_0})\; .
\end{equation}
Combining Eqs. (\ref{estimation1}) and (\ref{estimation2}), we get
\begin{equation}\label{estimationfinal}
A_{\vv{u}{0}:p_0,\vv{w}{0}:q_0} \leq  a_{\vv{u}{0}:p_0,\vv{w}{0}:q_0}
\leq B_{\vv{u}{0}:p_0,\vv{w}{0}:q_0}\; ,
\end{equation}
where 
\[
A_{\vv{u}{0}:p_0,\vv{w}{0}:q_0} = \max\left(- h(M_{\vv{u}{0}:p_0}) - h(M_{\vv{w}{0}:q_0}), \; h(M_{\vv{\mbox{-}w}{0}:p_0}) + h(M_{\vv{\mbox{-}u}{0}:q_0}) -2\right)
\]
and
\[
B_{\vv{u}{0}:p_0,\vv{w}{0}:q_0} = \min\left( 2-  h(M_{\vv{u}{0}:p_0}) - h(M_{\vv{w}{0}:q_0}), \; h(M_{\vv{\mbox{-}w}{0}:p_0}) + h(M_{\vv{\mbox{-}u}{0}:q_0})\right) \; .
\]
From the above expressions for $A_{\vv{u}{0}:p_0,\vv{w}{0}:q_0}$ and $B_{\vv{u}{0}:p_0,\vv{w}{0}:q_0}$, we obtain that $B_{\vv{u}{0}:p_0,\vv{w}{0}:q_0} - A_{\vv{u}{0}:p_0,\vv{w}{0}:q_0}\leq 2$, so there are at most three integers between these bounds. However, in many cases the bounds are much more narrow and there is only one integer between them.

Having reduced the number of possibilities for $a_{\vv{u}{0}:p_0,\vv{w}{0}:q_0}$, the remaining ones can be identified using a computer. For that propose, for each split function $h$, for each combination of $a_{\vv{u}{0}:p_0,\vv{w}{0}:q_0}$, $\lbrace\vv{u}{},\vv{w}{}\rbrace\in\CP,\; p,q\in \{1,2\}$ (satisfying Eq.\ (\ref{estimationfinal})), we used Eq.\ (\ref{evaluation}) to evaluate $\fff{q_1}{q_2}{q_3}{q_4}{q_5}{q_6}{q_7}$ and check whether the obtained value belongs to $Q$. Of course, the function $f$ is a local rule if and only if for any $q_1,\ldots,q_7\in Q$ the answer is positive. Table \ref{tab:finalresults} presents the results of this investigation described above.
Summing up all entries in this table, we get that there are exactly $5302$ three-dimensional three-state number-conserving CAs with the von Neumann neighborhood.

\begin{table}[h]
	\centering
    \begin{tabular}{|l|c|c|c|c|c|c|c|}
\hline
\diagbox{$h^{(2)}$}{$h^{(1)}$}  & $1000000$ & $0100000$ & $0010000$ & $0001000$ & $0000100$ & $0000010$ & $0000001$\\
    \hline
$2000000$ & 26 & 8 & 8 & 13 & 8 & 8 & 0 \\\hline
$0200000$ & 8 & 26 & 8 & 13 & 8 & 0 & 8 \\\hline
$0020000$ & 8 & 8 & 26 & 13 & 0 & 8 & 8 \\\hline
$0002000$ & 13 & 13 & 13 & 19 & 13 & 13 & 13 \\\hline
$0000200$ & 8 & 8 & 0 & 13 & 26 & 8 & 8 \\\hline
$0000020$ & 8 & 0 & 8 & 13 & 8 & 26 & 8 \\\hline
$0000002$ & 0 & 8 & 8 & 13 & 8 & 8 & 26 \\\hline
$1100000$ & 50 & 50 & 34 & 37 & 34 & 12 & 12 \\\hline
$1010000$ & 50 & 34 & 50 & 37 & 12 & 34 & 12 \\\hline
$1001000$ & 50 & 36 & 36 & 47 & 36 & 36 & 14 \\\hline
$1000100$ & 50 & 34 & 12 & 37 & 50 & 34 & 12 \\\hline
$1000010$ & 50 & 12 & 34 & 37 & 34 & 50 & 12 \\\hline
$1000001$ & 22 & 24 & 24 & 27 & 24 & 24 & 22 \\\hline
$0110000$ & 34 & 50 & 50 & 37 & 12 & 12 & 34 \\\hline
$0101000$ & 36 & 50 & 36 & 47 & 36 & 14 & 36 \\\hline
$0100100$ & 34 & 50 & 12 & 37 & 50 & 12 & 34 \\\hline
$0100010$ & 24 & 22 & 24 & 27 & 24 & 22 & 24 \\\hline
$0100001$ & 12 & 50 & 34 & 37 & 34 & 12 & 50 \\\hline
$0011000$ & 36 & 36 & 50 & 47 & 14 & 36 & 36 \\\hline
$0010100$ & 24 & 24 & 22 & 27 & 22 & 24 & 24 \\\hline
$0010010$ & 34 & 12 & 50 & 37 & 12 & 50 & 34 \\\hline
$0010001$ & 12 & 34 & 50 & 37 & 12 & 34 & 50 \\\hline
$0001100$ & 36 & 36 & 14 & 47 & 50 & 36 & 36 \\\hline
$0001010$ & 36 & 14 & 36 & 47 & 36 & 50 & 36 \\\hline
$0001001$ & 14 & 36 & 36 & 47 & 36 & 36 & 50 \\\hline
$0000110$ & 34 & 12 & 12 & 37 & 50 & 50 & 34 \\\hline
$0000101$ & 12 & 34 & 12 & 37 & 50 & 34 & 50 \\\hline
$0000011$ & 12 & 12 & 34 & 37 & 34 & 50 & 50 \\\hline
    \end{tabular}
 \caption{The numbers of three-dimensional three-state number-conserving CAs with the von Neumann neighborhood for each of 196 split functions $h$ (given by setting $h^{(1)}$ and $h^{(2)}$). }
  \label{tab:finalresults}
\end{table}
 
\section{Future work and open problems}
\label{sec:future}

Our new approach to study $d$-dimensional number-conserving CAs with the von Neumann neighborhood allows to revisit unresolved questions  in the field. 
First of all, future work will focus on determining whether Conjecture \ref{binary} is true or not. Since now we know all three-dimensional three-state number-conserving CAs, we will also be able to identify reversible ones -- in particular, to see whether there are some nontrivial ones. In~\cite{enum} we found that if the state set is $\{0,1\}$ or $\{0,1,2\}$, \emph{i.e.}, when we deal with binary or three-state CAs, there are only trivial number-conserving CAs that are reversible as well, namely, the identity and the shifts (in each of four possible directions). 
We conjecture that the same will hold for $d=3$. A natural extension of this problem is to determine whether the following general fact is true or not: three states are too few to enable the existence of  nontrivial reversible number-conserving CAs with the von Neumann neighborhood in any dimension.  

Although the decomposition theorem seems to have been developed for higher dimensions, it is useful for $d=1$ as well. For example, existing tools allow to find all one-dimensional reversible number-conserving CAs with four states, while also a list of 20 reversible number-conserving CAs with five states is given \cite{imai2018radius}, but it is unknown whether the list is complete. The decomposition theorem allows us to find all number-conserving five-state CAs and identify the reversible ones. 

The basic idea of the split-and-perturb decomposition of the local rule of a number-conserving CA, presented here for a cubic grid and the von Neumann neighborhood, can be used in other settings as well, for example, for two-dimensional triangular or hexagonal grids. The difficulty then, however, lies in describing the space of perturbations. This will be the subject of our future work.


\bibliographystyle{spmpsci}
\bibliography{decomposition}

\end{document}